# Quantum key secure communication protocol via enhanced superdense coding

Mario Mastriani

**Summary** In the last decades, Quantum Cryptography has become one of the most important branches of Quantum Communications with a particular projection over the future Quantum Internet. It is precisely in Quantum Cryptography where two techniques stand out above all others: Quantum key Distribution (QKD), and Quantum Secure Direct Communication (QSDC). The first one has four loopholes relating to the exposure of the key at all points in the communication system, while the second has clear practical implementation problems in all its variants currently in use. Here, we present an alternative to QKD and QSDC techniques called quantum key secure communication (QKSC) protocol with a successful implementation on two free access quantum platforms.

ORCID Id: 0000-0002-5627-3935

Mario Mastriani: mmastria@fiu.edu

School of Computing & Information Sciences, Florida International University, 11200 S.W. 8th Street, Miami, FL 33199, USA

## 1 Introduction

Currently, information security implies, one way or another, the use of techniques based on principles of quantum mechanics. Below we will mention the most relevant of all of them:

*Quantum key distribution* (QKD) protocols. They base their entire operation on the use of polarized[1,2] or entangled photons[3,4], being the manipulation of the key its flank of greater exposure to the attack of an eavesdropper, which compromises their efficiency in real life applications. This eventual weakness comes in the form of four security loopholes which are related to: 1) the exposure of the key in the quantum channel (QCh) or distribution channel, 2) the transfer of the key from said channel to the sender terminals, 3) the transfer of the key from the same channel to the receiver terminals, and 4) the free availability of the public channel through which the encrypted message or ciphertext travels. These loopholes are commonly ignored in both communication terminals, which generates a serious security problem due to the daily advance of eavesdropping techniques.

*Quantum no-key* protocol[5,6]. This protocol is divided into three stages, where each one has its own key, consisting of unitary transforms applied to the message transferred at each stage. Its main differences with a QKD protocol, e.g. Bennett-Brassard (BB84)[1], consists in that here both the keys and the message always remain quantum, and that at each stage the flow of communication has an opposite direction to the previous one.

*Deterministic secure quantum communication* (DSQC)[7,8] protocol. This quantum protocol uses a key to encrypt each classical bit of a deterministic message thanks to two photons. In the same way, as in the case of quantum cryptography techniques based on QKD, DSQC protocol can work with entangled or polarized photons, using a unidirectional quantum channel, and a classical channel[7,8]. Besides, as a sine qua non condition, in DSQC the information must be preprocessed by photons before encoding it in order to avoid unnecessary exposure of parts of the message in the channel in the presence of an eventual eavesdropper. Whether or not such presence is detected, the same protocol will be the vehicle to send the key by which the message will be decrypted.

*Quantum secure direct communication* (QSDC) protocol[9-19]. It was created in 2000 by Prof. Gui-Lu Long from the Department of Physics of Tsinghua University. This protocol eliminates the four loopholes mentioned above in relation to the key management in QKD, therefore, it constitutes the greatest expression of quantum cryptography, being QSDC the best quantum technique for secure communications currently in use. In fact, QSDC exploits the attributes of the entanglement[20-22] for the development of long-distance quantum communication[23-26] like no other. Consequently, it can directly transmit secure information through quantum channels without keys[10]. In particular, the simplified version of QSDC, presented in this work, is extremely ductile when we try to extend their range via any kind of quantum repeaters[27-31], whether we use lines of optical fiber or if necessary, satellite quantum repeaters[32,33].

*Post-quantum cryptography* (PQC)[34,35]. They are a set of techniques completely separate from those mentioned above, which are composed by a series of tools applied from classical computers that allow immunizing both computers and communications systems from attacks by future quantum computers. Among the most representative PQC techniques, we can mention hash tree or Merkle tree[36], and elliptic curves[37], where the latter provides a similar level of encryption to the Rivest-Shamir-Adleman (RSA) algorithm[38] but with keys comparatively much smaller in size.

Consequently, after having studied the pros and cons of the two most used techniques in Quantum Cryptography: QKD[1-4], and QSDC[9-19], we present here an alternative to both techniques without their respective defects called Quantum Key Secure Communication (QKSC) protocol, which is extremely robust and easy to implement in both optical[39] and superconductor[40-43] platforms. However, we resort here to the 16-qubits Melbourne IBM quantum processor[40], and Quirk simulator[44], without losing generality in the number of qubits, which theoretically can be extended into an infinite number, or at the origin of the implementation platform[40-43].

Showing up next, the main tool necessary to implement the QKSC protocol, i.e., an enhanced version of the Super-dense Coding protocol[45,46] is developed in Sec. 2. QKSC protocol is explained in Sec. 3. Implementations of the novel on the 16-qubits Melbourne IBM quantum processor[40], and Quirk simulator[44] are presented in Sec. 4 with a complete section about the analysis of the outcomes. Finally, Sec. 5 provides the conclusions.

## 2 Enhanced Super-dense Coding protocol

All the keys used in the cryptographic protocols mentioned in Sec. 1 are expressed in bits or in their quantum counterparts, which arise from the two poles of the Bloch's sphere[46-48] and which are called Computational Basis States (CBS). These CBS can be expressed in several ways:

$$North\ pole = Spin\ up = |0\rangle = \begin{bmatrix} 1 \\ 0 \end{bmatrix}, \text{ and} \tag{1}$$

$$South\ pole = Spin\ down = |1\rangle = \begin{bmatrix} 0 \\ 1 \end{bmatrix}. \tag{2}$$

QKD protocols such as Bennett and Brassard's 1984 (BB84)[1] also use qubits of the type:

$$|+\rangle = H|0\rangle = \frac{1}{\sqrt{2}}(|0\rangle + |1\rangle), \text{ and } |-\rangle = H|1\rangle = \frac{1}{\sqrt{2}}(|0\rangle - |1\rangle), \tag{3}$$

where $H$ is the Hadamard gate[46-48]. QKSC uses classic bits and the famous Bell bases[46-48]:

$$|\Phi^+\rangle = \frac{1}{\sqrt{2}}(|00\rangle + |11\rangle) \equiv |\beta_{00}\rangle, \qquad |\Phi^-\rangle = \frac{1}{\sqrt{2}}(|00\rangle - |11\rangle) \equiv |\beta_{10}\rangle,$$

$$|\Psi^+\rangle = \frac{1}{\sqrt{2}}(|01\rangle + |10\rangle) \equiv |\beta_{01}\rangle, \text{ and } |\Psi^-\rangle = \frac{1}{\sqrt{2}}(|01\rangle - |10\rangle) \equiv |\beta_{11}\rangle. \tag{4}$$

In fact, the type of qubits used to be transmitted securely thanks to the QKSC protocol will be exclusively the bases of Eq.(4) and their extension to Greenberger–Horne–Zeilinger (GHZ) states[46]. Specifically, the QKSC protocol is based on the use of an enhanced version of the Super-dense Coding technique[45,46], which will be analyzed below along with the original protocol. Figure 1 shows three versions of such protocol, where:

a) represents the original protocol[45], such that:
   - in $t_0$, we have two ancillas $|0\rangle$ in q[0] and q[1], and both classical bits {$b_0$, $b_1$} to be transmitted in a secure way, where the simple lines represent quantum wire, i.e., lines that transport qubits, while the double lines have to do with classic wires, i.e., lines that transport classic bits,
   - in $t_1$, both Bell states $|\beta_{00}\rangle = \frac{1}{\sqrt{2}}(|00\rangle + |11\rangle)$ are available in q[0] and q[1],
   - between $t_2$ and $t_1$, both classic bits {$b_0$, $b_1$} are incorporated to the quantum circuit via a pair of Controlled-Pauli's gates (Controlled-X, CNOT or CX; and Controlled-Z, or CZ)[46-48] applied on the qubit q[0], and
   - between $t_3$ and $t_2$, an unitary transform is applied on qubits q[0] and q[1] which is formed by a CNOT and a Hadamard (H) gate[46], besides both quantum measurements[49,50] are carried out, in order to obtain the transmitted classic bits {$b_0$, $b_1$} from the recovered CBS allocated in q[0] and q[1].

b) represents the enhanced protocol, such that:
   - in $t_0$, we have two ancillas $|0\rangle$ in q[0] and q[1], and both classic bits {$b_0$, $b_1$} to be transmitted in a secure way, exactly the same as the original protocol case,
   - in $t_1$, both Bell states $|\beta_{00}\rangle = \frac{1}{\sqrt{2}}(|00\rangle + |11\rangle)$ are available in q[0] and q[1], in the same way as in the original protocol,
   - between $t_2$ and $t_1$, the difference among the original protocol and the novel appears, since the CZ gate is applied to qubit q[1], while the CNOT gate is applied to qubit q[0], and
   - between $t_3$ and $t_2$, the same unitary transform based on a CNOT and a Hadamard (H) gate[46] is applied on qubits q[0] and q[1]. Besides both quantum measurements[49,50] are carried out, in order to obtain the transmitted classic bits {$b_0$, $b_1$} from the recovered CBS allocated in q[0] and q[1]. However, both in the novel and in the original protocol the outputs lose the original order of the classic bits transmitted when they are recovered, as shown in Fig. 1(b). So we turn to the next modified version of the improved protocol.

c) represents the modified enhanced protocol, such that:
   - in $t_0$, it is identical to the two previous versions,
   - in $t_1$, it is identical to the two previous versions,
   - between $t_2$ and $t_1$, it is identical to the last version,
   - between $t_3$ and $t_2$ a beam-splitter, formed by a CNOT gate in q[0] and a Hadamard gate in q[1], is flipped respect to the last version, in order to obtain the transmitted classic bits {$b_0$, $b_1$} from the recovered CBS allocated in q[0] and q[1] in the correct order. See Fig. 1(c).

In this study, we will use the version (c) of Fig. 1, because it constitutes a primal stage of the QKSC protocol, which will be described in the next section. In practice, we will use symbols or samples (of a signal) coded with several classic bits (not just two), therefore, it is important to have a previous example of the aforementioned enhanced protocol for a greater number of bits to be transmitted safely, which is not possible with the original super-dense coding[45,46] protocol, since it only supports two. Figure 2 shows a generalization of the enhanced super-dense coding protocol of Fig. 1(c) for 4 classic bits {$b_0$, $b_1$, $b_2$, $b_3$} to be transmitted. As we can see, this configuration is a natural extension of the 2 classic bits version, with a single point to highlight, which is between $t_2$ and $t_1$. The passage between the protocol of Fig. 1(c) for 2 bits and that of Fig. 2 for 4 bits consists in increasing the number of CNOT gates used as the number of classic bits to be transmitted increases, while the CZ gate is only applied to the most significant qubit in both versions, in fact, q[1] for Fig. 1(c) and q[3] for Fig. 2.

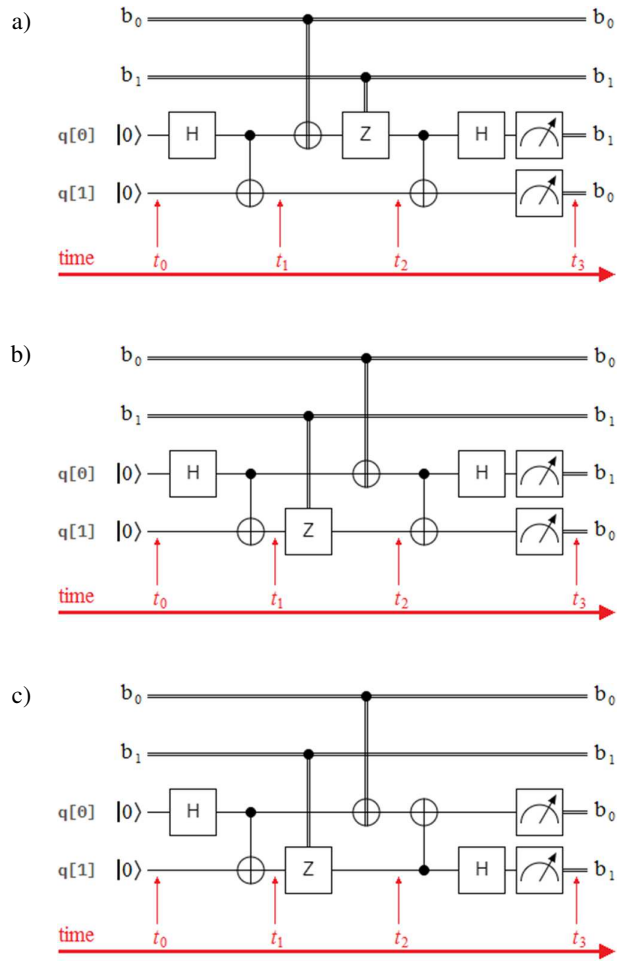

FIG. 1: Three versions of the super-dense coding protocol, where, a) is the original protocol[45,46], b) is the enhanced version, and c) is the enhanced version with a modification necessary to recover the transmitted classic bits in the correct order.

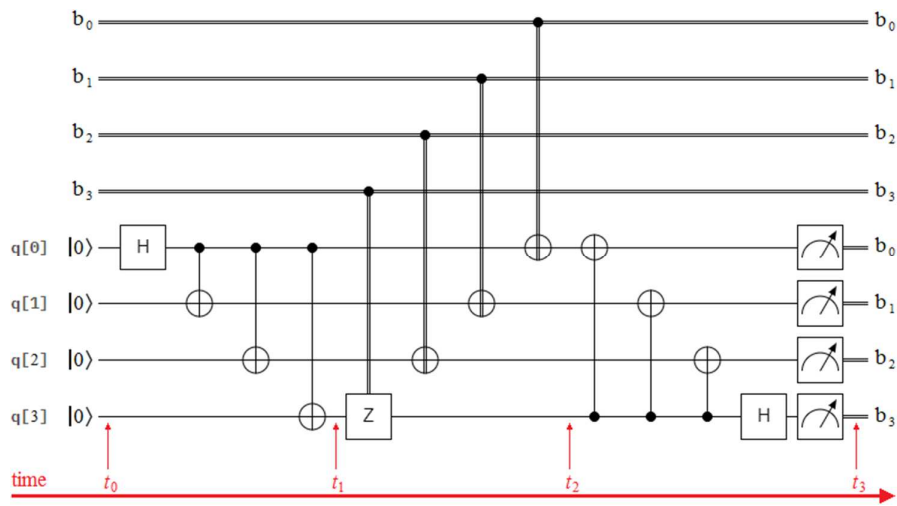

FIG. 2: Generalization of the enhanced super-dense coding protocol of Fig. 1(c) for 4 classic bits $\{b_0, b_1, b_2, b_3\}$ to be transmitted.

Table I summarizes the time slot for Figures 1 and 2, where, $I^{N \times N}$ means identity matrix of $N$-by-$N$ elements, $\otimes$ is the Kronecker product[46], and $|GHZ_4\rangle = (|0000\rangle + |1111\rangle)/\sqrt{2}$ is a Greenberger-Horne-Zeilinger state for four entangled particles[46-48]. Finally, Fig. 2 as a natural extension of Fig. 1(c), where the number of classical bits to be teleported can be extended unlimitedly.

Table I: Time slot of Fig. 1 and 2.

| time | Figure | | | |
|---|---|---|---|---|
| | 1(a) | 1(b) | 1(c) | 2 |
| 0 | $\{b_0, b_1\} \in \{0,1\}$ $q[0] = q[1] = |0\rangle$ | $\{b_0, b_1\} \in \{0,1\}$ $q[0] = q[1] = |0\rangle$ | $\{b_0, b_1\} \in \{0,1\}$ $q[0] = q[1] = |0\rangle$ | $\{b_0, b_1, b_2, b_3\} \in \{0,1\}$ $q[0] = q[1] = q[2] = q[3] = |0\rangle$ |
| 1 | $\{b_0, b_1\} \in \{0,1\}$ $q[0] = q[1] = |\beta_{00}\rangle$ | $\{b_0, b_1\} \in \{0,1\}$ $q[0] = q[1] = |\beta_{00}\rangle$ | $\{b_0, b_1\} \in \{0,1\}$ $q[0] = q[1] = |\beta_{00}\rangle$ | $\{b_0, b_1, b_2, b_3\} \in \{0,1\}$ $q[0] = q[1] = q[2] = q[3] = |GHZ_4\rangle$ |
| 2 | $\{b_0, b_1\} \in \{0,1\}$ $\{q[0], q[1]\} =$ $(Z^{b_1} X^{b_0} \otimes I^{2 \times 2})|\beta_{00}\rangle$ | $\{b_0, b_1\} \in \{0,1\}$ $\{q[0], q[1]\} =$ $(X^{b_0} \otimes I^{2 \times 2})(I^{2 \times 2} \otimes Z^{b_1})|\beta_{00}\rangle$ | $\{b_0, b_1\} \in \{0,1\}$ $\{q[0], q[1]\} =$ $(X^{b_0} \otimes I^{2 \times 2})(I^{2 \times 2} \otimes Z^{b_1})|\beta_{00}\rangle$ | $\{b_0, b_1, b_2, b_3\} \in \{0,1\}$ $\{q[0], q[1], q[2], q[3]\} =$ $(X^{b_0} \otimes I^{8 \times 8})(I^{2 \times 2} \otimes X^{b_1} \otimes I^{4 \times 4})$ $(I^{4 \times 4} \otimes X^{b_2} \otimes I^{2 \times 2})(I^{8 \times 8} \otimes Z^{b_3})|GHZ_4\rangle$ |
| 3 | $q[0] = b_0$, $q[1] = b_1$ | $q[0] = b_0$, $q[1] = b_1$ | $q[0] = b_0$, $q[1] = b_1$ | $q[0] = b_0$, $q[1] = b_1$, $q[2] = b_2$, $q[3] = b_3$ |

### 3 Quantum key secure communication (QKSC) protocol

All cryptographic configurations based on QKD protocols[1-4], like that of Fig. 3(a), work with three types of channels:

- Quantum channel (QCh), in red in Fig. 3(a), which results from the distribution of polarized[1,2] or entangled[3,4] photons, which are generated by an emitter (which we will call Alice), and sent through from this quantum channel (QCh) to a receiver (which we will call Bob).

- Public or classical channel (PCh), in black in Fig. 3(a), which is used to complete tasks related to the key distribution, e.g., key sifting and distillation[51] in the case of the BB84 protocol[1]. The sifting task essentially consists of a process where the bits are eliminated, both from the sender and the receiver, which do not have an exact correlation between them. So when this task is done, the sender and the receiver share a key of the same length known as the sifted key[52]. An eavesdropper cannot reconstruct the key from the information exposed in the channel during this task, since the errors related to said key have to do with the intervention of the eavesdropper and not with the key itself. As a consequence of this, a second process called key distillation is launched. Basically, the process known as key distillation is composed of two stages: a) in the first stage all the errors of the key are corrected by means of a classic procedure[52], while b) in the second one the key is conveniently reduced by means of a procedure known as privacy amplification so that the eavesdropper does not have the necessary information.

- Data or insecure channel (DCh), in light blue in Fig. 3(a), is the weakest link in the chain and is the natural medium through where the encrypted message or ciphertext travels, which arises from applying the distributed key to the unencrypted original message or plaintext. As in the previous case, this channel is also public and classic in nature.

Instead, quantum secure direct communication (QSDC) protocols[9-19], Fig.3(b), uses a classical channel for authentication to verify a correct mutual authentication between Alice and Bob, and a quantum channel with two branches (forward and backward) to complete the transmission of the message. Formally, QSDC does not work with a ciphertext like QKD, however, we allowed ourselves to call the coded message that travels thanks to the quantum channel ciphertext in order to establish a better comparison between QSDC and QKD of Figs. 3(a) and 3(b), respectively. In fact, there are excellent previous examples of QSDC protocols using GHZ states[53,54], however all of them without a key.

Like all quantum secure direct communication (QSDC) protocols[9-19], the configuration of the QKSC protocol of Fig. 3(c) proposed in this work uses an authenticated public channel or service channel in order to regularly verify the integrity of the transmitted data between Alice and Bob[55], and a quantum channel QCh as a transport agent for the encrypted message, for which, prima facie, no key is generated or distributed, and the message is encoded in Bell bases like those seen in Eq.(4) which are generated and encoded in the sender (Alice) and travel through the quantum channel QCh to the receiver (Bob) where they are decoded, as part of the process to recover the message, which in this case will be symbols of a text or samples of a signal. At this point, the most noticeable difference between QSDC and QKSC is that the former uses a two-branch quantum channel, while the latter works with a single-branch quantum channel, i.e., the forward branch. In fact, the mere fact of eliminating a branch of the quantum channel remarkably simplifies the practical implementation of QKSC versus QSDC, and this is precisely the second most newsworthy difference between them: the simplicity of implementation of QKSC versus QSDC, both in superconducting quantum computers and on optical circuits. Specifically, QKSC can interchangeably work with symbols coded, for example, in American Standard Code for Information Interchange (ASCII) code[56], where every symbol is coded in 8 bits; or work with signal samples coded in a pulse-code modulation (PCM) codification[57] with $n$ bits-per-sample (bps). In all cases these are binary and classic bits to be transmitted from Alice (sender) to Bob (receiver) using the QKSC protocol.

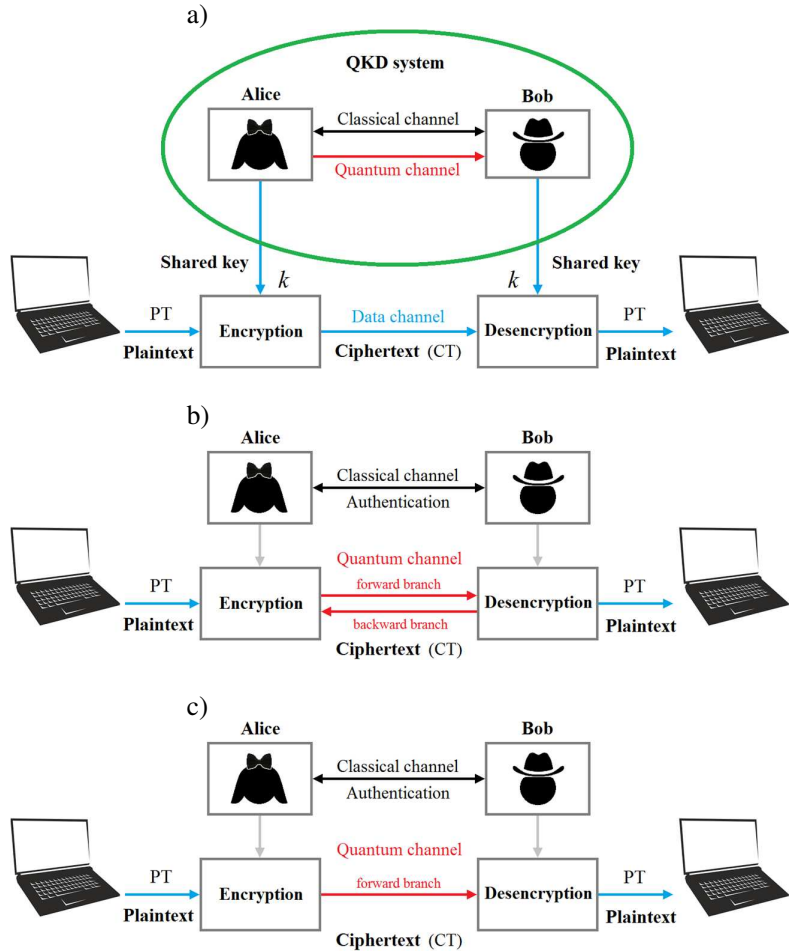

FIG. 3: Schematic comparison between QKD, QSDC, and QKSC protocols. a) QKD system with a classical channel (black), and a quantum channel (red) for the key distribution; while the encrypted message travels via another classical channel (blue) called data channel. b) QSDC protocol with a classical channel for authentication (black), and a quantum channel (red) with two branches: forward, and backward. c) QKSC protocol with an identical configuration like that of QSDC, with only one exception: its quantum channel only needs the forward branch.

For transmissions over long distances, using lines of optical fiber or satellite links, QKSC will require, like any other quantum cryptography protocol, the use of quantum repeaters between Alice (sender) and Bob (receiver), as can be seen in Fig. 4, which shows the block diagram of the QKSC protocol, where Alice incorporates $n$ ancillas of the type $|0\rangle$ and the symbol (or sample) coded in $n$ classic bits.

At the output of this block, there is a quantum channel (QCh) which consists of a single wire because this block incorporates a multiplexer at its output. The diagram continues with a number of quantum repeaters, which depends on the type of link used, i.e., by lines of optical fiber or satellite link and the distance to be covered with the communication system. Each quantum repeater incorporates a multiplexer-demultiplexer pair, hence these blocks exclusively work with a single wire. Finally, Bob receives the coded information from the only wire that constitutes the quantum channel (QCh) and after a measurement he recovers the classic $n$ bits of the transmitted symbol (or sample) B.

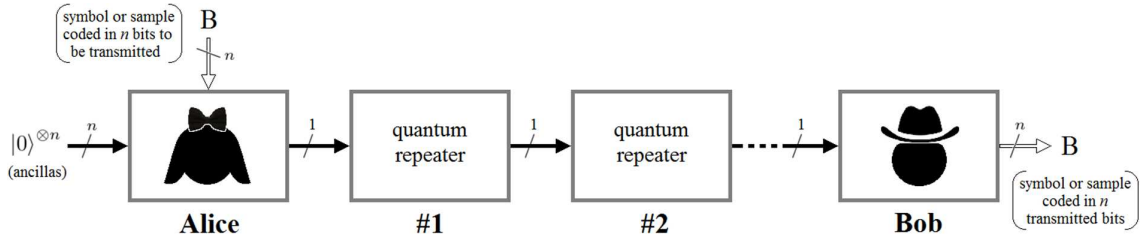

FIG. 4: Block diagram of the QKSC protocol, where Alice (sender) works with $n$ ancillas of the $|0\rangle$ type, and every symbol B (or sample) is coded in $n$ classic bits. At the output of Alice, there is a single wire due to the internal use of a multiplexer. Each quantum repeater has its own multiplexer-demultiplexer pair. Bob (receiver) has a demultiplexer after which it performs a measurement that allows the transmitted symbol (or sample) B to be recovered.

In principle, this protocol uses a single channel, the QCh, however, faced with the possibility of losing photons due to:

- problems in their transmission through lines of optical fiber or through spatial transmission via satellite links, or by
- attacks by an eavesdropper of the photon-number splitting attack[58,59] type,

it may be necessary to use an authenticated public channel or a service channel in order to regularly verify the integrity of the transmitted data. However, since QKSC works essentially on the basis of entanglement, any method of evaluating an attack based on the measurement of entanglement can be used, in the same way as QKD protocols based on entangled photons[3,4]. On the other hand, the current use of single photon sources can aid in the detection of certain types of attacks, such as, photon-number splitting attack[58,59]. Later, in the *security analysis* subsection, we will develop our own policy to preserve the integrity of the information transmitted against any type of attack.

Figure 5 shows the internal constitution of Alice's and Bob's modules as an example of a symbol B coded in four bits $\{b_0, b_1, b_2, b_3\}$. Alice's module begins, from left to right of Fig. 5(a), with four ancillas $|0\rangle$ in qubits $\{q[0], q[1], q[2], q[3]\}$, which feed a generating source of a $GHZ_4$ state[20-22]:

$$GHZ_4 = \frac{1}{\sqrt{2}}\left(|0000\rangle + |1111\rangle\right). \tag{5}$$

Until a few years ago, it was experimentally challenging to construct optical multiplexers (MX) and demultiplexers (DX) in the single photon level, however, today there are several successful examples of their implementations[60-64]. However, QKSC is preserved by a dynamic key present at both ends of the communication system, which allows it to successfully use MX/DX pairs at the multi-photon level[65], and still be immune to attacks by an eavesdropper of the photon-number splitting attack[58,59] type.

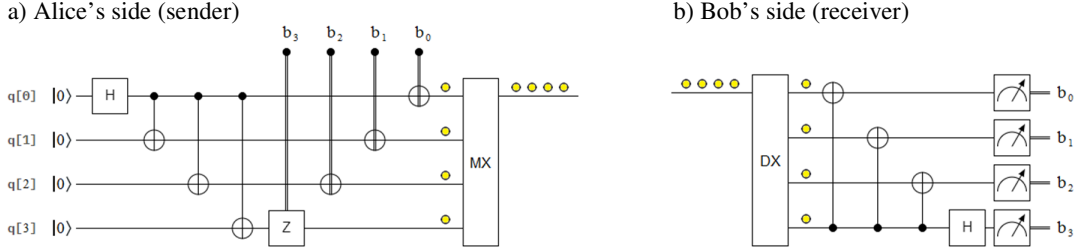

FIG. 5: Internal representation of Alice's and Bob's modules, as an example of four bits to be transmitted, where a) consists of four ancillas of the $|0\rangle$ type in {q[3], q[2], q[1], q[0]}, an H (Hadamard) and three CNOTs gates that generate a $GHZ_4$ state, one Controlled-Z and three CNOTs gates that allow incorporating the four bits {$b_3$, $b_2$, $b_1$, $b_0$} of symbol (or sample) in this module, and one multiplexer (MX) which is a submodule that converts the parallel set of photons (yellow circles) to serial; and b) consists of a demultiplexer (DX), i.e., a submodule that converts the serial set of photons to parallel, with three flipped CNOTs and one H gates, and four quantum measurement constituted in practice by single photon detectors, thanks to which the four transmitted bits {$b_3$, $b_2$, $b_1$, $b_0$} of a symbol or sample B, are recovered.

The four outputs of the $GHZ_4$ state source are intercepted for three CNOT gates, and a Controlled-Z (CZ) gate in their more significant qubit q[3]. This encodes each wire of the $GHZ_4$ state with its corresponding bit of a B symbol. This will be seen in detail in a later section called *implementation on two quantum platforms*. Then, a multiplexer (MX) converts a parallel flow of photons to a series one, which are represented as little yellow circles. On the other hand, Fig. 5(b) shows Bob's module, which begins with a demultiplexer (DX) that converts a series flow of photons to a parallel one. Then, Bob applies a measurement module like that of Fig. 2 between times $t_2$ and $t_3$. This makes it possible to recover the four bits {$b_0$, $b_1$, $b_2$, $b_3$} of a transmitted symbol B.

Figure 6(a) shows a parallel configuration of a quantum repeater that uses teleportation swapping (TS) in each photon, see Fig. 6(b). TS is based on a technique called entanglement swapping[66-72], which is the most frequently used in quantum communications[23-26] for the construction of quantum repeaters[27-31]. In Fig. 6(a), the photons' stream enters to the demultiplexer (DX), which allows the photons to separate in a parallel configuration. Then, each photon enters to a TS module, which extends the range of the link. We can accumulate several TSs, one after the other, within each quantum repeater, or use repeaters of a single TS per qubit and use several quantum repeaters along the link in order to multiply the link range. Finally, all the photons that come out of each TS go to a multiplexer (MX) which aligns them in the quantum channel (QCh). In Fig. 6(b), it is possible to see in detail the interior of each TS. Besides, it is also possible to use a configuration that uses a single TS for all four photons at the same time, i.e., a serial configuration like that of Fig. 6(c)[27-31], which is applied directly on the quantum channel (QCh) without the need to separate each photon as in the configuration of Fig. 6(a) thanks to a multiplexer-demultiplexer pair.

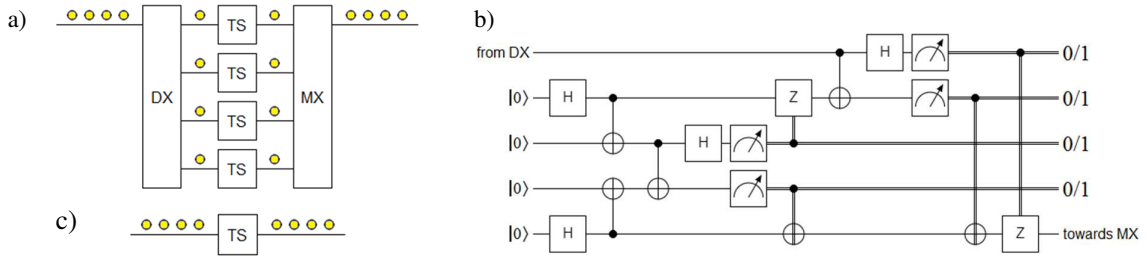

FIG. 6: Quantum repeaters based on teleportation swapping (TS): a) functional diagram for the example of Fig. 5, where each wire requires its own TS after the demultiplexer (DX), then all photons go to the multiplexer (MX) to continue the transmission; b) detail of the internal circuitry of each TS, where a simple line represents a quantum wire, i.e., a line that transports qubits, while a double line means classic wire, i.e., a line that transports classic bits; and c) serial configuration of the quantum repeater for the four photons at the same time.

Regarding which is the best configuration of quantum repeater: if the parallel of Fig. 6(a), or the series of Fig. 6(c); this is still a matter of investigation, since it is exclusively a topic linked to the implementation of the repeaters, which depends on the communication system design factors associated with the distance to be covered and the type of repeaters selected, which are based on lines of optical fiber, or via space link. In the case that the distribution of entangled photons takes place thanks to lines of optical fiber, it is necessary to use repeaters every 50 km due to the losses in the fiber, an attenuation in the material, and a propagation speed (*v*) equal to 2/3 of the speed of light (*c*), while satellite repeaters can cover greater distances at a speed $v = c$, with less attenuation and losses than in the case of optical fiber except for relative environmental aspects to the ground-sky link, i.e., clouds that can disrupt the distribution of entangled photons.

QKSC can interchangeably use any of the quantum repeater configurations mentioned in Fig 6. Besides, it can even use a more direct family which is presented here as an alternative version, and that is shown in Fig 7. The configurations of Fig 7 work according to a completely different criterion than those in Fig. 6. These techniques base their operation on two main characteristics:
   a) old photons which come in and new ones which come out, and
   b) they only work forward, i.e., there are no quantum repeaters in the middle of the quantum channel (QCh) distributing entangled photons back and forth in order to use the transitivity of entanglement swapping[66-72].

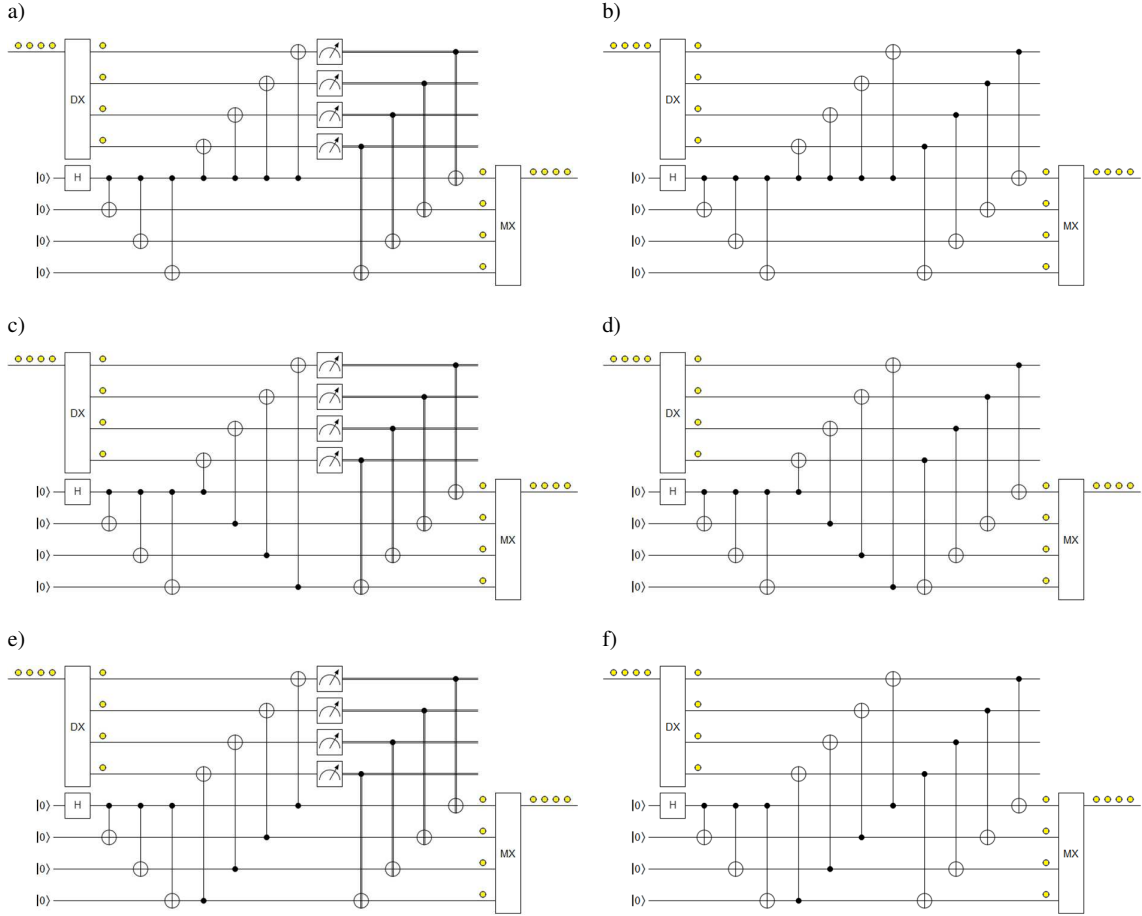

FIG. 7: Proposed quantum repeater. The photons that come from the demultiplexer (DX) interact with the GHZ$_4$ states and after a measurement process new photons are obtained with the same attributes with which they went out from Alice's module. Versions (a, c, and e) are perfectly interchangeable between them, while (b, d, and f) are their equivalent and simplified versions, respectively. Finally, a multiplexer converts the organization of the resulting photons from a parallel to a series configuration, so that they continue their journey on the quantum channel.

The configurations shown in Fig. 7 are exclusively parallel, where the photons that come from the demultiplexer (DX), and at the same time from the quantum channel (QCh), interact with GHZ$_4$ states and after a measurement process new photons are obtained with the same attributes with which they went out from Alice's module. Versions of Fig. 7(a, c, and e) are perfectly interchangeable between them, while Fig. 7(b, d, and f) are their equivalent and simplified versions, respectively. In fact, in a later section called *implementation on two quantum platforms*, we will use the simplified version of Fig. 7(b) in order to simplify implementations on both platforms. Finally, a multiplexer converts the organization of the resulting photons from a parallel to a series configuration, so that they continue their journey on the quantum channel.

To summarize, the quantum repeater scheme proposed in Fig. 7 represents a relay mechanism of old photons for new ones, exclusively forward, in which, as in the case of the quantum repeaters[27-31] commonly used in practice such as those from Fig. 6, it will require quantum memories[73] for its optical implementation in order to compensate for differences in the arrival time of each photon.

**QKSC protocol.** Based on Fig. 4, we can define this protocol as a procedure for the case where the quantum channel is a line of optical fiber, where we can analyze the status of the quantum channel at the beginning and end of each module:

$$t = 0, QCh(t) = |0\rangle^{\otimes n},$$
$$t = t+1, QCh(t) = O_{Alice}\left[B(t),|0\rangle^{\otimes n}\right],$$
$$t = t + 1,$$
$$r = 0,$$
*while* (*range*-50 km) > *r*,
{
$$QCh(t) = O_{QR}\left[QCh(t-1)\right],$$
  $r = r + 50$ km,
  $t = t + 1,$
} *end while*,
$$B(t) \leftarrow QCh(t) = O_{Bob}\left[QCh(t-1)\right].$$

Where:
  *range* is the distance between sender and receiver,
  $QCh(t)$ is the quantum channel status in each time,
  $B(t)$ is the symbol to be transmitted in each time,
  $n$ is the size of the symbol B, i.e., the number of bits to be transmitted,
  $O_{Alice}$ is the Alice's oracle,
  subscript QR means quantum repeaters,
  $O_{QR}$ is the QR's oracle,
  $O_{Bob}$ is the Bob's oracle,
  $t$ is the time,
  $r$ is the accumulator associated with the number of quantum repeaters, and
  50 km represents the distance between modules in case of working with lines of optical fiber.

As we can see, the procedure described above is simple, however, we need to take into account an important detail, which consists of the inside of Alice's module, in which case, we have to generate GHZ$_n$ states. If *n* is a large number, the implementation of Alice's module may seem a complicated task in practice. However, through a configuration as in the work of Prof. Pan[74] group, which generates GHZ$_4$ states, we can extend this to a general case of GHZ$_{2^k}$ (*k* is the level of splitting) states from a single-photon source and an optical demultiplexer. This implies a pulse laser with a central wavelength around 893 nm excites the quantum dot resonantly through a fiber-coupled confocal microscope system, polarizing beam-splitter, half- and quarter-wave plates, Pockels cell, and mirrors for its implementation. In Fig. 8(a), a quantum circuitry interpretation of this configuration is presented, while Figures 8(b), 8(c), 8(d), and 8(e), are extensions of the mentioned configuration for the case of $GHZ_8$ is developed through four equivalent quantum circuits.

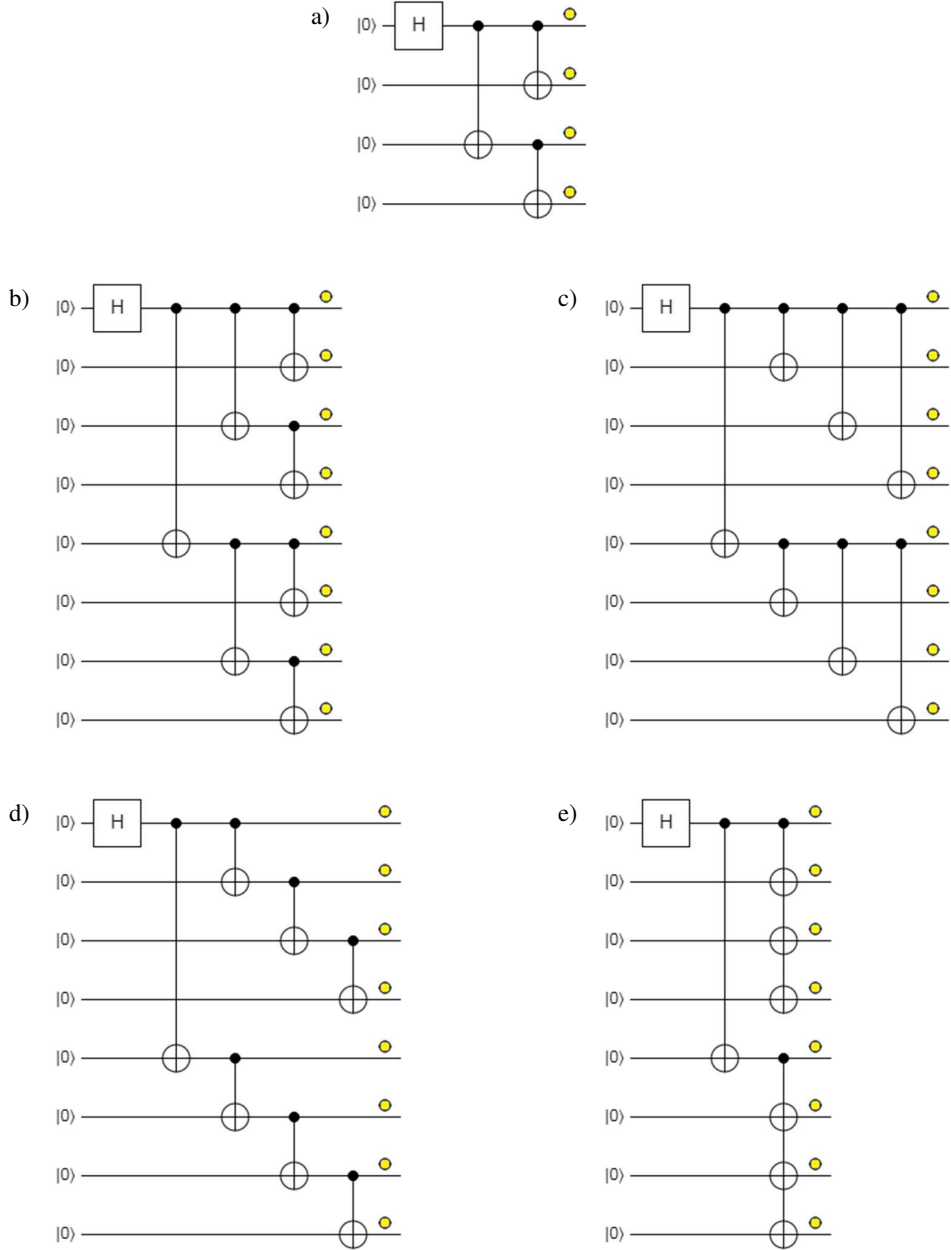

FIG. 8: Circuitry interpretation of $GHZ_{2^k}$ ($k$ is the level of splitting) states generation from a single-photon source and an optical demultiplexer, where: a) is the representation of the example of $GHZ_4$ presented in the paper of Prof. Pan[74] group, while from b) to e) are representing, with the same criterion, four equivalent versions of $GHZ_8$.

Finally, all this simplicity could be transferred to the work of an eavesdropper, even working with sources that generate the minimum necessary amount of photons, which apparently would protect the configuration of a photon-number splitting attack[58,59], for which we must do a more exhaustive analysis of security.

**Security analysis.** Since QKSC bases its entire operation on entangled photons, it is possible to consider measures to prevent the negative effects of the intervention of eavesdroppers of the type:

a) entanglement measurement as in the case of Ekert 91 (E91) protocol[3],

b) the use of the exact, minimum and necessary number of photons in order to prevent interventions like photon-number splitting attacks[58,59].

However, if we consider the transmission losses either in lines of optical fibers, or via a space link, we will find that in order to prevent an attack we expose ourselves to the loss of critical photons, thus compromising the integrity of the message, therefore, QKSC contemplates the possibility of using a multiple-photon source accompanied by a key K, that is shared between Alice (sender) and Bob (receiver), and that should only be transmitted the first time via:

i) a manual procedure, i.e., in a pre-arranged way, or

ii) through the use of a QKD protocol, either polarized[1,2], or entangled[3,4] photons.

QKSC protocol itself transmits its next keys, starting, inclusive, with the second key. In Fig. 9, a generic block diagram using a key on Alice (sender) and Bob (receiver) is established.

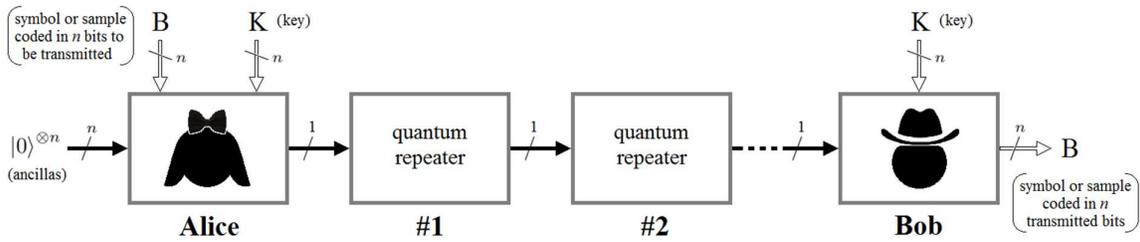

FIG. 9: Block diagram of the QKSC protocol, similar to that of Fig. 4 but with a key K, where only Alice's (sender) and Bob's (receiver) modules receive the key.

As we can see, scheme of Fig. 9 is similar to that of Fig. 4, where the intervention of the key K only affects Alice's and Bob's modules, and not to quantum repeaters.

Figure 10 shows the internal representation of Alice's and Bob's modules with the intervention of a key K {$k_3, k_2, k_1, k_0$}, as an example of four bits to be transmitted {$b_3, b_2, b_1, b_0$}, where the only difference with the modules of Fig. 5 consists in the incorporation of the key K {$k_3, k_2, k_1, k_0$} on Alice's and Bob's qubits {q[3], q[2], q[1], q[0]}. This incorporation is via three CNOTs, and one Controlled-Z gate on the most significant qubit q[3] of both modules. Moreover, in the case of Alice's module the key K is incorporated after the symbol B and before the multiplexer (MX), in the case of Bob's module, the key K is incorporated after the demultiplexer (DX) and before the measurement that Bob must take to retrieve the transmitted symbol B.

The example of Fig. 10 is extensive to symbols and keys of any size, with the sole exception of the number of qubits supported by the host platform. In this work, we will implement the QKSC protocol in the section called *implementation on two quantum platforms*, where said platforms will be Quirk simulator[44], of 16 qubits, and a 16-qubits IBM Q processor[40], called Melbourne, which has the highest number of qubits of free access in the IBM Q series non-Premium quantum processor family. Therefore, these platforms will condition our ability to test the protocol with large symbols and keys, e.g., for 1024, 2048, 4096 or more bits. Furthermore, in real implementations carried out through optical circuits[39], the use of quantum memories[73] should be considered, both in the sender and in the receiver. For example, on the receiver's side it is necessary to keep the current key until the symbol to be decrypted arrives.

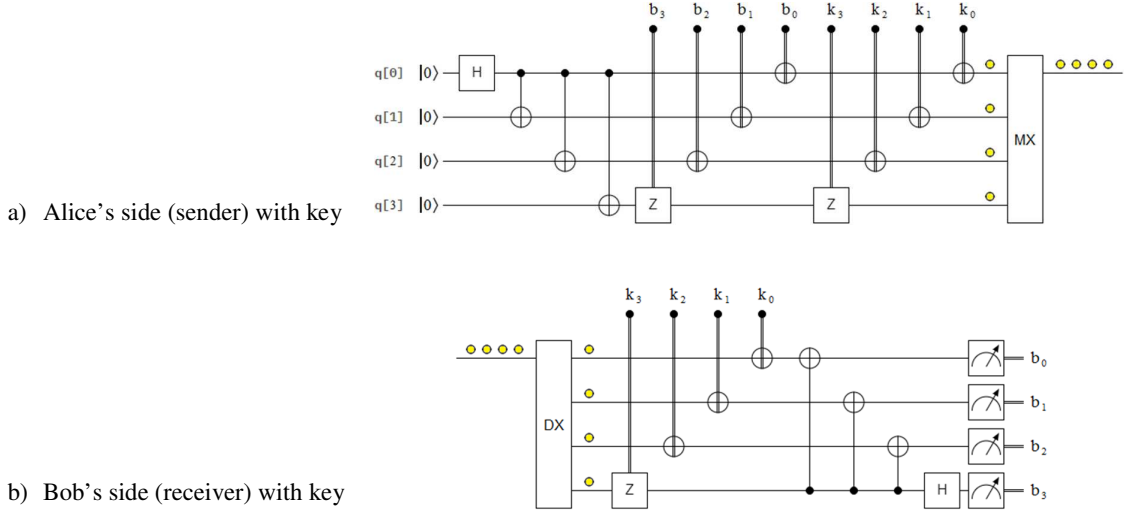

a) Alice's side (sender) with key

b) Bob's side (receiver) with key

FIG. 10: Internal representation of Alice's and Bob's modules, as an example of four bits to be transmitted, where the only difference with the modules of Fig. 5 consists in the incorporation of the key K {$k_3$, $k_2$, $k_1$, $k_0$} on Alice's and Bob's qubits {q[3], q[2], q[1], q[0]}. This incorporation is via three CNOTs, and one Controlled-Z gate on the most significant qubit q[3] of both modules. While in the case of Alice's module the key is incorporated after the symbol B and before the multiplexer (MX), in the case of Bob's module, the key is incorporated after the demultiplexer (DX) and before the measurement.

Finally, the optical implementation of the QKSC protocol, i.e., Alice's and Bob's modules, and the details about the constitution of optical quantum repeaters based on Bell state measurement (BSM) modules, quantum memories, and photon detectors are outside the scope of this study, however, it is important to take into account recent experiments carried out by Prof. Pan's group, in which optical quantum repeaters are implemented without the use of quantum memories[75].

**QKSC protocol with a static key.** Based on Fig. 9, we can define a version of this protocol as a procedure that uses a key for the case where the quantum channel is a line of optical fiber. We can analyze the status of the quantum channel at the beginning and end of each module as in the keyless case:

$$t = 0, QCh(t) = |0\rangle^{\otimes n},$$
$$t = t+1, QCh(t) = O_{Alice}\left[B(t), K, |0\rangle^{\otimes n}\right],$$
$t = t + 1,$
$r = 0,$
while (range-50 km) > r,
{
$$QCh(t) = O_{QR}\left[QCh(t-1)\right],$$
$r = r + 50$ km,
$t = t + 1,$
} end while,
$$B(t) \leftarrow QCh(t) = O_{Bob}\left[QCh(t-1), K\right].$$

Where K is an n bits shared key, which is previously distributed between Alice and Bob thanks to any QKD protocol[1-4], any other QSDC protocol[9-19], or manually, i.e., pre-arranged. After that, any new innovation of the key will be via QKSC itself using the last key to protect the transmission. Finally, in the procedure described above, the only modules impacted by the key are those of Alice and Bob, since the quantum repeaters are identical to those used in the keyless version of this procedure.

**QKSC protocol with a dynamic key.** Based again on Fig. 9, we can define a new version of this protocol as a procedure that uses a dynamic key for the case where the quantum channel is a line of optical fiber. Then, we need to define what criteria we will base on to dynamically evolve the key, for which we highlight the best way to do it among an innumerable number of them. The first key K(0) is generated thanks to a QRNG, distributed manually, via any QKD protocol[1-4], or via any other QSDC protocol[9-19], while from the second one onwards, the current key K(t) is calculated from an XOR operation between the previous symbol B(t-1) and the previous key K(t-1), i.e., K(t) = B(t-1) $\veebar$ K(t-1), that is, the complete scheme is a conditional of the type:

    if *t = 0*,
       K(t) = K(0),
    else
       K(t) = B(t-1) $\veebar$ K(t-1),
    end if.

Reviewing the four dynamic key proposals mentioned above, we can see that we have gone from the highest to the lowest in the exposure of the key in the channel, since in the first option each key generated must be distributed and therefore exposed in the channel, while in the case of the last option, the exposure is minimal because the first key is distributed only, while from that moment on the key is dynamically and independently updated in a synchronized and simultaneous way in both Alice's and Bob's side without the need of any transmission. Therefore, based on the fourth option, we will modify the QKSC protocol with the inclusion of the dynamic key:

$$t = 0, QCh(t) = |0\rangle^{\otimes n},$$
$$t = t+1, QCh(t) = O_{Alice}\left[B(t), K(t), |0\rangle^{\otimes n}\right],$$
$t = t + 1$,
$r = 0$,
*while* (range-50 km) > *r*,
{
    $$QCh(t) = O_{QR}\left[QCh(t-1)\right],$$
    $r = r + 50$ km,
    $t = t + 1$,
} *end while*,
$$B(t) \leftarrow QCh(t) = O_{Bob}\left[QCh(t-1), K(t-1)\right].$$

Moreover, since experiments indicate that the key does not introduce noise, we can use it permanently, even in a broader security context. Finally, for both static and dynamic keys (ignoring the presence of MX/DX) the Oracle and outcomes of Alice and Bob will be:

$$O_{Alice} = O_{Alice,2}\, O_{Alice,1}$$
$$O_{Alice,1} = \left(X^{b_0} \otimes I^{8\times 8}\right)\left(I^{2\times 2} \otimes X^{b_1} \otimes I^{4\times 4}\right)\left(I^{4\times 4} \otimes X^{b_2} \otimes I^{2\times 2}\right)\left(I^{8\times 8} \otimes Z^{b_3}\right)$$
$$O_{Alice,2} = \left(X^{k_0} \otimes I^{8\times 8}\right)\left(I^{2\times 2} \otimes X^{k_1} \otimes I^{4\times 4}\right)\left(I^{4\times 4} \otimes X^{k_2} \otimes I^{2\times 2}\right)\left(I^{8\times 8} \otimes Z^{k_3}\right)$$
$$\{q[0], q[1], q[2], q[3]\} = QCh = O_{Alice}|GHZ_4\rangle$$

$$O_{Bob} = O_{Bob,2}\, O_{Bob,1}$$
$$O_{Bob,1} = \left(X^{k_0} \otimes I^{8\times 8}\right)\left(I^{2\times 2} \otimes X^{k_1} \otimes I^{4\times 4}\right)\left(I^{4\times 4} \otimes X^{k_2} \otimes I^{2\times 2}\right)\left(I^{8\times 8} \otimes Z^{k_3}\right)$$
$$O_{Bob,2} = \left(I^{8\times 8} \otimes H\right)\left(I^{4\times 4} \otimes CX_{flipped}\right)\left(I^{4\times 4} \otimes SW\right)\left(I^{2\times 2} \otimes CX_{flipped} \otimes I^{2\times 2}\right)\left(I^{4\times 4} \otimes SW\right)\left(I^{2\times 2} \otimes SW \otimes I^{2\times 2}\right)\left(CX_{flipped} \otimes I^{4\times 4}\right)$$
$$\{q[0], q[1], q[2], q[3]\} = O_{Bob}\, QCh = \{b_0, b_1, b_2, b_3\}$$

where $CX_{flipped}$ is a flipped CNOT gate (control and target qubit are reversed), and SW is a SWAP gate.

## 4 Implementation on two quantum platforms

In this section, we will carry out a series of experiments without and with the use of a key previously distributed between Alice's and Bob's modules. In all cases, we will work with three-bit symbols for reasons of limitation in the number of qubits supported by the two platforms on which we will carry out the aforementioned implementations: Quirk simulator[44], and 16-qubits Melbourne IBM Q processor[40]. Quirk was chosen for being the most powerful non-commercial platform in terms of its visual expression and the number of metrics it uses to evaluate the outcomes of the experiments, while in the case of Melbourne, in addition to being the non-Premium processor with the greatest number of qubits of the IBM Q[40] family, which implies greater decoherence and errors compared to the series of Premium processors from the same company, it is a physical machine which will allow us to analyze the performance of the novel in a real environment. On the other hand, in all the experiments, we will compare the results obtained with Melbourne and the IBM Q simulator[40], which delivers identical results to Qiskit[40], in order to check the deterioration introduced in Melbourne by decoherence[50], and different types of errors like bit-flip, phase-flip, and bit-phase-flip[76].

Finally, all the implementations in this section will have the following points in common:

- it is not possible to implement both the multiplexer (MX) and the demultiplexer (DX), for reasons related to their construction and operation, so their implementations will be exclusively in charge of an optical version,
- the symbol $B = \{b_2 = 0, b_1 = 1, b_0 = 1\}$ to be transmitted will be the same in all experiments, and
- the GHZ base will be the same in all cases, that is,

$$GHZ_3 = \frac{1}{\sqrt{2}}(|000\rangle + |111\rangle) \cdot \qquad (6)$$

**Keyless with only one quantum repeater:** Figure 11 shows an implementation of QKSC protocol on Quirk simulator[44] transmitting the mentioned symbol B. Alice's side first stage consists in the symbol preparation where we can build CBS as |1> thanks to NOT gates (i.e., Pauli's inverter X) after the ancillas |0>. Then, we use quantum measurements to obtain the bits of the symbol B from their respective CBS of Eqs.(1 and 2). The second stage is reserved for the $GHZ_3$ state generation, like that of Eq.(6). The third and last stage of Alice's side inserts the elements of symbol $B = \{b_2 = 0, b_1 = 1, b_0 = 1\}$ as control bits in two CNOT gates and one Controlled-Z gate, while the $GHZ_3$ states enter the same gates but at their target qubits.

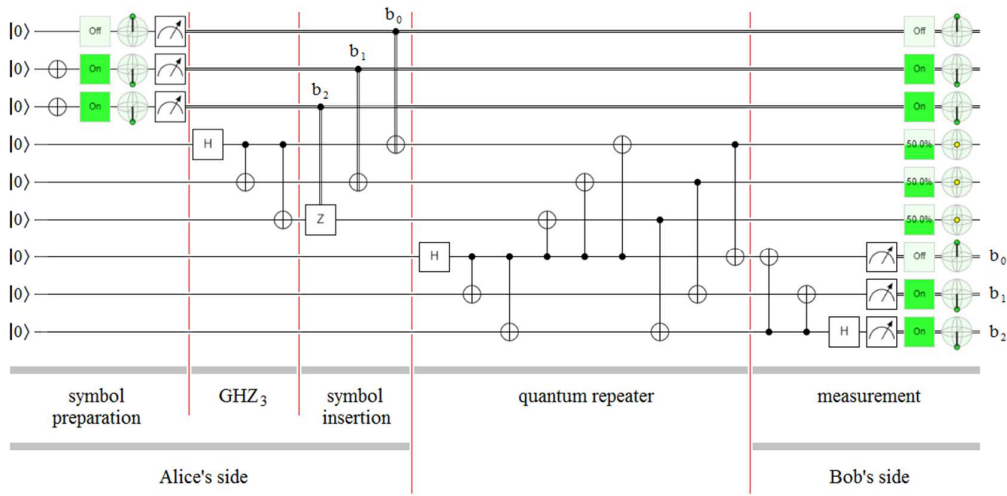

FIG. 11: QKSC implementation on Quirk simulator, keyless, and with only one quantum repeater.

The second module belongs to the only quantum repeater implemented in this experiment, which corresponds to the simplified version of Fig. 7(b) but for three qubits and not including the multiplexer (MX) or the demultiplexer (DX) used in said figure.

Finally, in the third module Bob makes the measurement after two flipped CNOTs and one Hadamard (H) gates, obtaining the same bits that were generated in Alice's first module. We must remember that Quirk simulator is an ideal environment without decoherence or noise, which only allows us to evaluate the logical feasibility of the protocol. Therefore, for a more realistic evaluation about the operation of the QKSC protocol on a physical machine we must resort to a platform like 16-qubits Melbourne IBM Q processor[40], where Fig. 12 shows its connectivity map, with the single-qubit U2 and CNOT error rates in detail.

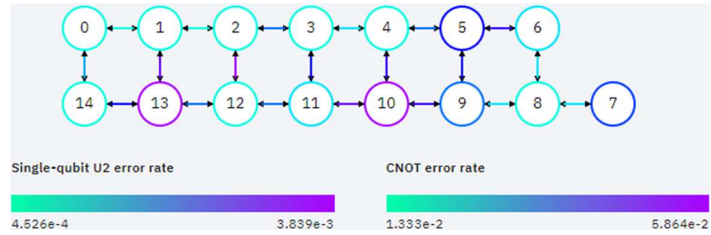

FIG. 12: Connectivity map of the 16-qubits Melbourne IBM Q processor[40], with the single-qubit U2 and CNOT error rates in detail.

In Fig. 13, an implementation of the QKSC protocol on the IBM Q platform is presented, where the fundamental differences with the implementation on Quirk simulator are the following:

- We have eliminated the measurements of the first stage of Alice's side, which converted CBS into their respective bits, so as not to dilate the histograms of Fig. 15, which corresponds to the execution on Melbourne processor, so as not to complicate the reading of their outcomes, and

- we implement the Controlled-Z gate of the most significant qubit of the third stage of Alice's module thanks to one CNOT gate, and two H gates (before and after the target qubit of the CNOT gate).

As in Quirk, the metrics to the right of Fig. 13, i.e., circles in gray and/or blue, show that the upper three qubits (to be transmitted) and the lower three (transmitted) coincide, which indicates they have been transmitted successfully.

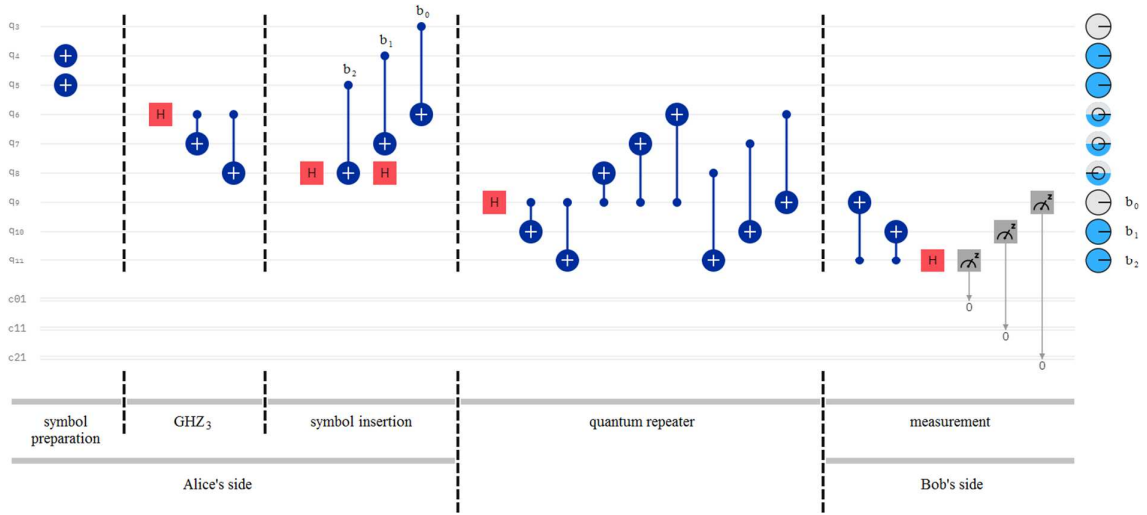

FIG. 13: QKSC implementation on IBM Q, keyless, and with only one quantum repeater.

Figures 14 and 15 show the outcomes of Fig. 13 when its circuits is ran on the IBM Q simulator and 16-qubits Melbourne quantum processor, respectively. In both cases, with 8192 shots and fairshare as the run mode. However, while Fig. 14 shows 100% of probability in $b_2b_1b_0$ = 110, Fig. 15 has 71.241% of probability in $b_2b_1b_0$ = MM0, and 28.759% of probability in $b_2b_1b_0$ = MM1, where M is a meta-symbol that represents 0 and 1, at the same time. This important dispersion in the outcomes of the 16-qubits Melbourne IBM Q processor compared to those obtained with the IBM Q simulator is a direct consequence of decoherence[50] and the aforementioned flip errors[75] present in Melbourne processor.

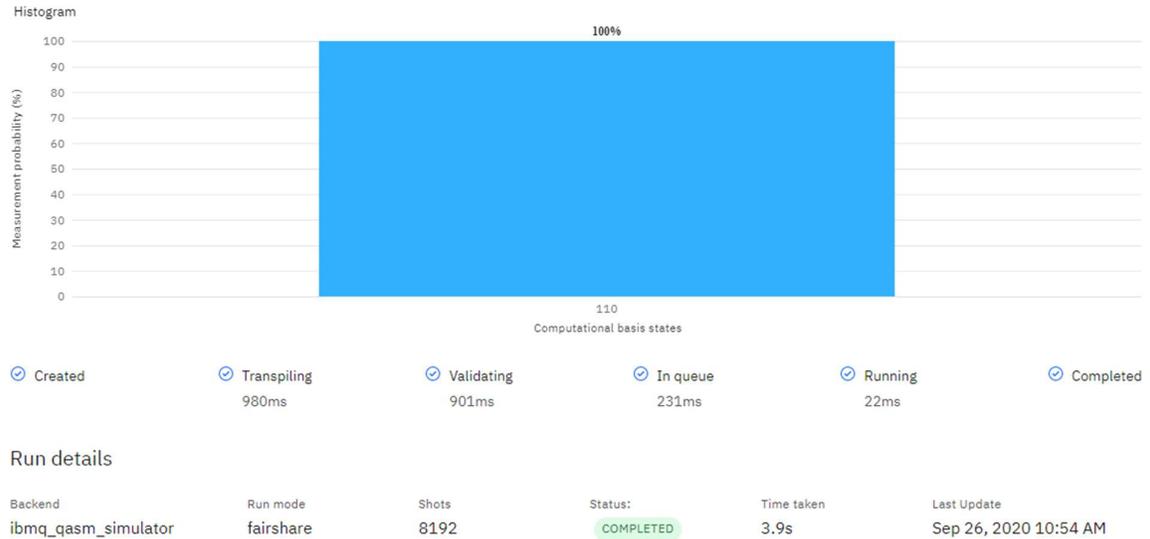

FIG. 14: IBM Q simulator. Upper figure: Histogram (measurement probabilities) in terms of computational basis states, with 100% of probability in $b_2b_1b_0$ = 110, which evidences an absolute coincidence with the metrics of Fig. 13. Lower figure: Run details of simulator execution, with 8192 shots, and fairshare run mode.

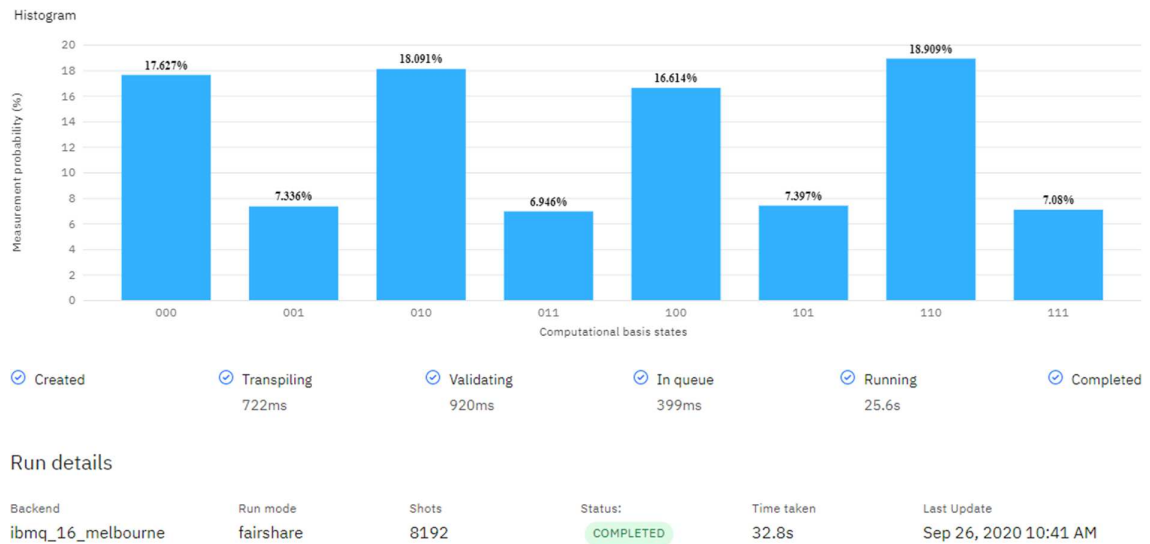

FIG. 15: IBM Q Melbourne processor. Upper figure: Histogram (measurement probabilities) in terms of computational basis states, with (17.627+18.091+16.614+18.909)% = 71.241% of probability in $b_2b_1b_0$ = MM0, and (7.336+6.946+7.397+7.08)% = 28.759% of probability in $b_2b_1b_0$ = MM1, where M is a meta-symbol that represents 0 and 1, at the same time. The difference of almost 29% between the simulator and Melbourne processor shows decoherence of the latter for this experiment. Lower figure: Run details of simulator execution, with 8192 shots, and fairshare run mode.

**Keyless, with two quantum repeaters:** Figures 16 and 17 show the implementation of the QKSC protocol on Quirk simulator, and IBM Q, respectively, without a key, and with two quantum repeaters. In both implementations, the coincidence between the bits to be transmitted, i.e., those higher qubits corresponding to Alice's first stage, and those received in the lower qubits on Bob's side, is total. In Quirk, the verification is automatic thanks to its clear visual metrics, while in IBM Q the circles on Bob's side mean: $b_2 b_1 b_0 \equiv$ blue-blue-gray $\equiv 110$, which are the correct outcomes.

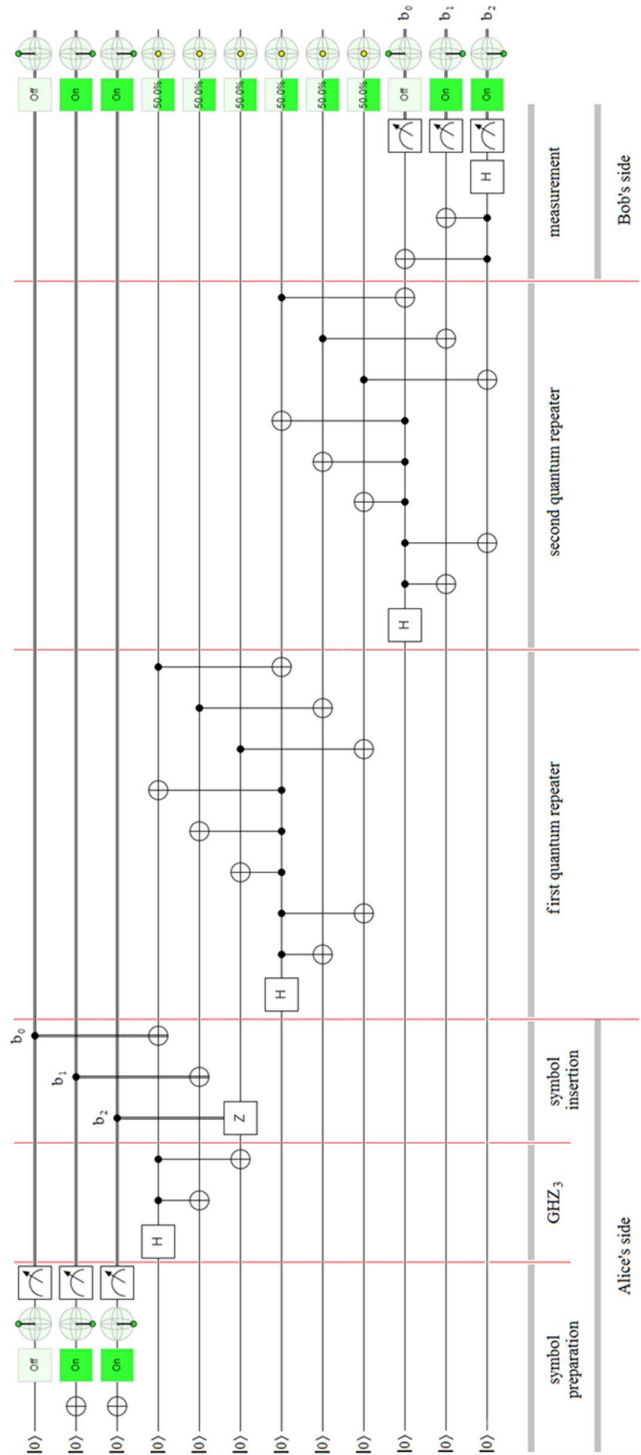

FIG. 16: QKSC implementation on Quirk simulator, keyless, and with two quantum repeaters.

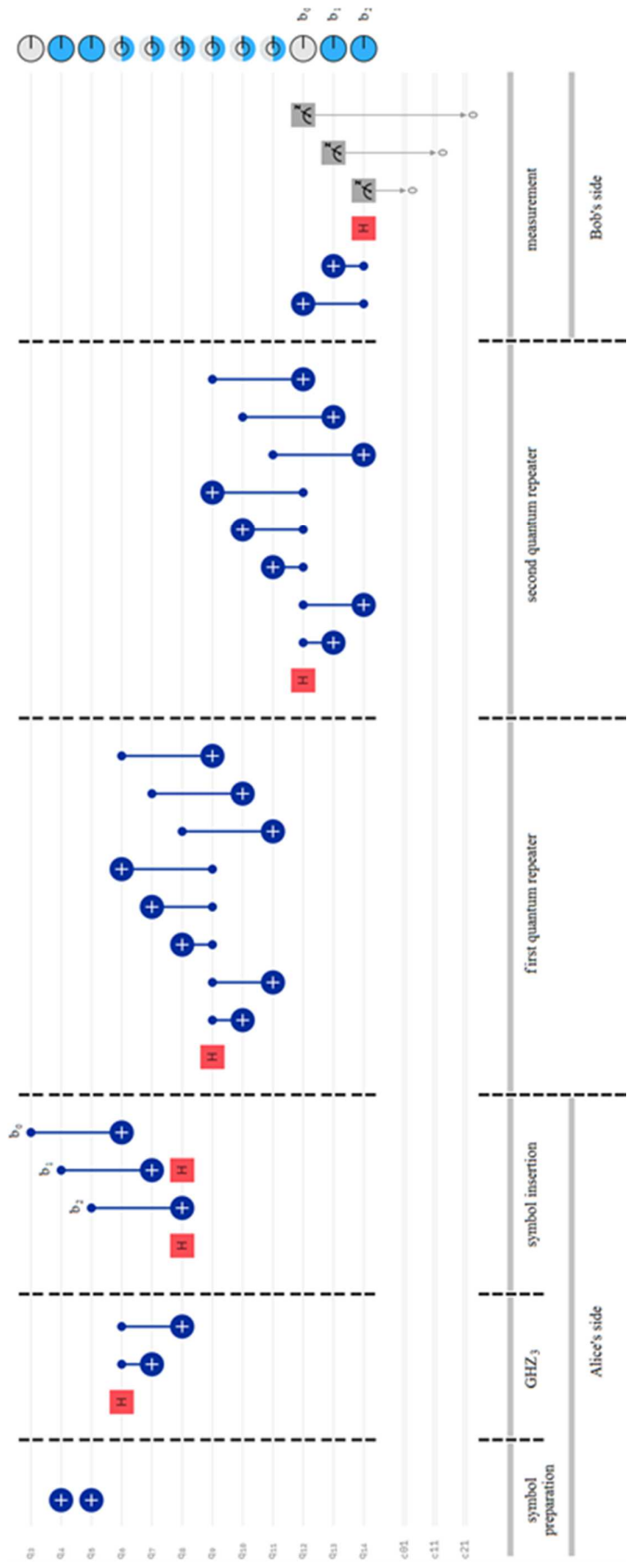

FIG. 17: QKSC implementation on IBM Q, keyless, and with two quantum repeaters.

As in the previous experiment, with a single quantum repeater, Fig. 18 shows the results presented by the IBM Q simulator, which are perfect with a 100% probability at $b_2b_1b_0 = 110$, and 0% for the rest of CBS. The problem arises when we analyze the outcomes delivered by the 16-qubit Melbourne IBM Q processor, which corresponds to less than 60% effectiveness, i.e., 59.497%, with more than 40% error. This increase in error with respect to the previous experiment has to do with the number of CNOT gates incorporated when adding the second quantum repeater, which adds bit flip errors[76], deteriorating the quality of outcomes.

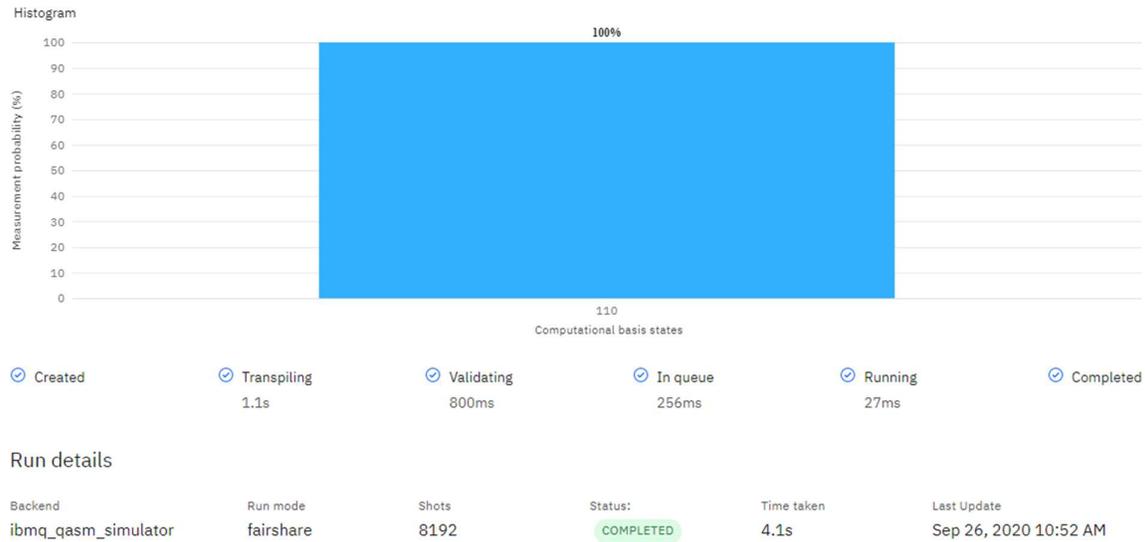

FIG. 18: IBM Q simulator. Upper figure: Histogram (measurement probabilities) in terms of computational basis states, with 100% of probability in $b_2b_1b_0 = 110$, which evidences an absolute coincidence with the metrics of Fig. 13. Lower figure: Run details of simulator execution, with 8192 shots, and fairshare run mode.

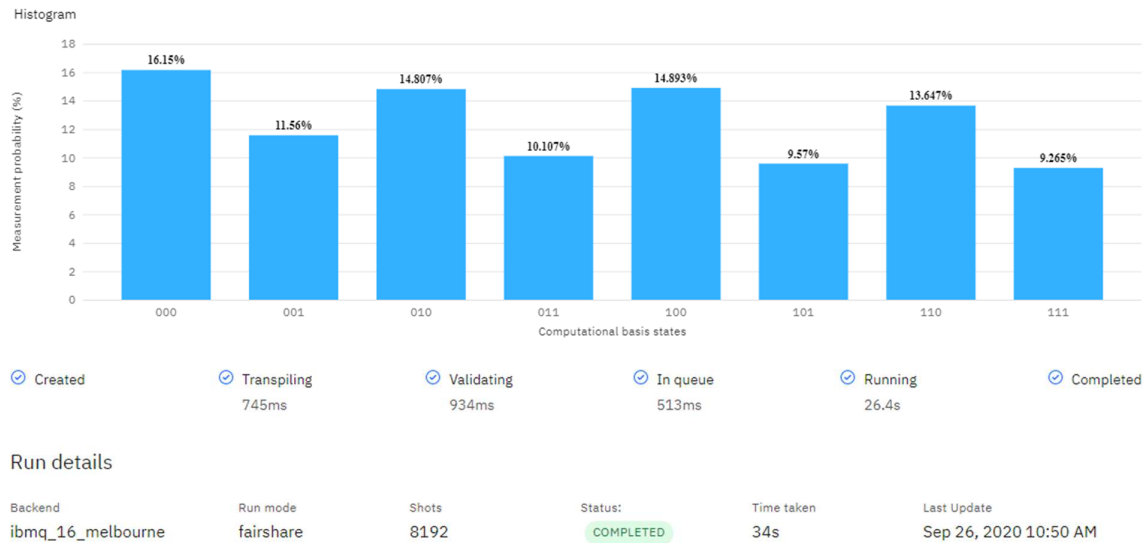

FIG. 19: IBM Q Melbourne processor. Upper figure: Histogram (measurement probabilities) in terms of computational basis states, with $(16.15+14.807+14.893+13.647)\% = 59.497\%$ of probability in $b_2b_1b_0 = MM0$, and $(11.56+10.107+9.57+9.265)\% = 40.503\%$ of probability in $b_2b_1b_0 = MM1$, where M is a meta-symbol that represents 0 and 1, at the same time. The difference is greater than 40% between the simulator and Melbourne processor showing decoherence of the latter for this experiment. Lower figure: Run details of simulator execution, with 8192 shots, and fairshare run mode.

**Key, and only one quantum repeater:** In Figs. 20 and 21 the implementations of the QKSC protocol are represented with a key and one quantum repeater only, for Quirk simulator[44], and IBM Q[40], respectively, where the symbol B is the same used in the last experiments, i.e., $b_2b_1b_0 = 110$, while the key K will be $k_2k_1k_0 = 101$. As in the previous cases, the metrics of both platforms indicate the success of the transmission, with 100% effectiveness. However, we must evaluate the impact of decoherence and flip type errors thanks to the intervention of more gates of the CNOT and Controlled-Z type in the Alice's (sender), and Bob's (receiver) modules on a physical platform like 16-qubits Melbourne IBM Q.

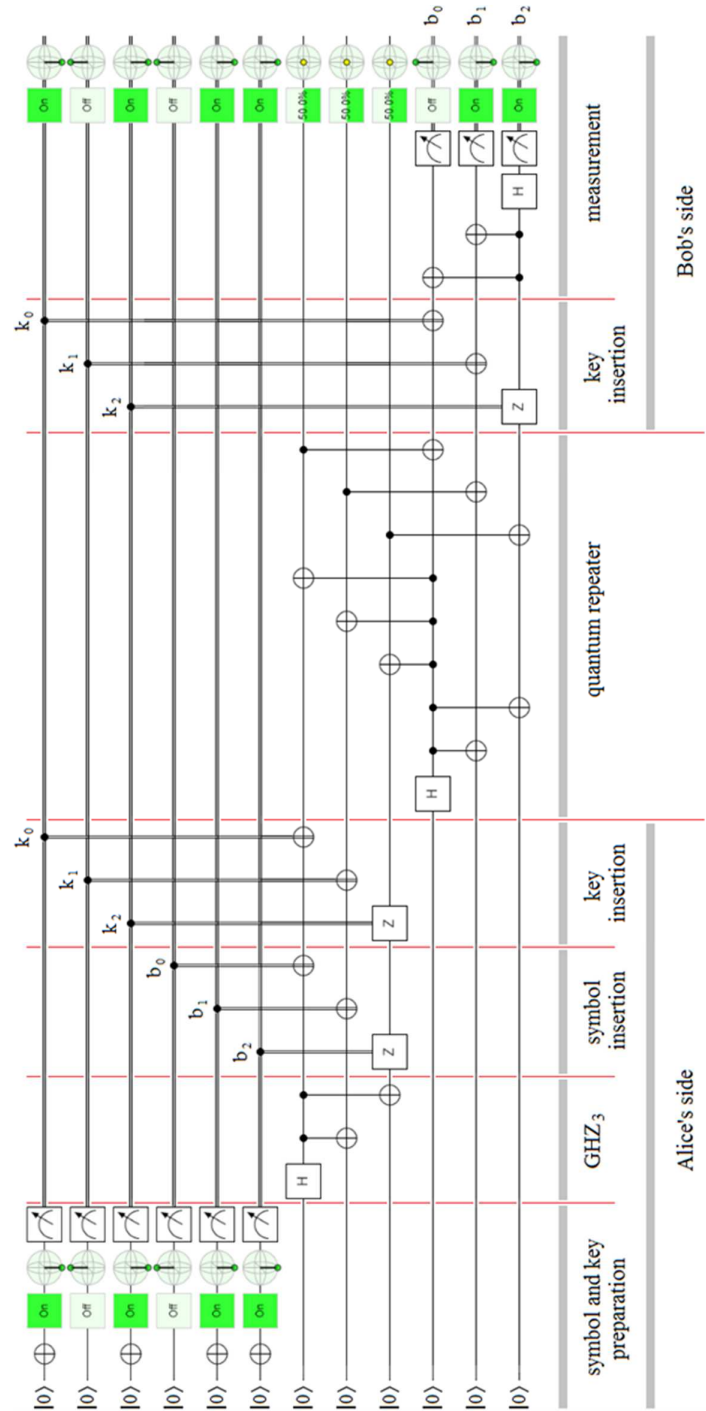

FIG. 20: QKSC implementation on Quirk simulator, with key, and only one quantum repeater.

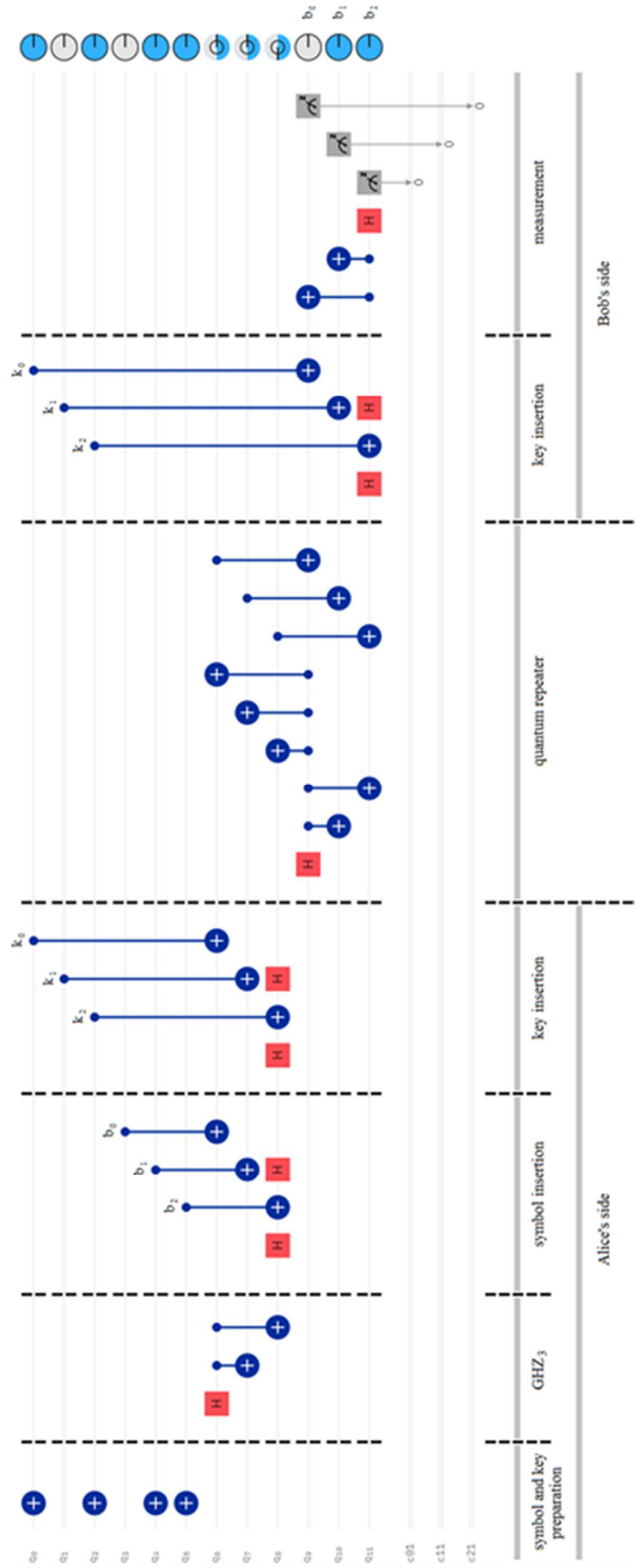

FIG. 21: QKSC implementation on IBM Q, with key, and only one quantum repeater.

In Fig. 22, it is possible to see an identical performance to those obtained in the implementations of Figures 14 and 18 for the keyless cases with one and two quantum repeaters, respectively. On the other hand, Fig. 23 shows the outcomes obtained in the IBM Q Melbourne processor, in which the impact of decoherence and flip errors is evident. However, the values of Fig. 23 resemble those of Fig. 15 (keyless with only one quantum repeater), being much higher in performance than those of Fig. 19 (keyless with two quantum repeaters). This establishes the suspicion that the key does not seem to be responsible for the drop in the performance, but rather the quantum repeaters, for which we must carry out a final experiment with a key and two quantum repeaters to confirm this hypothesis.

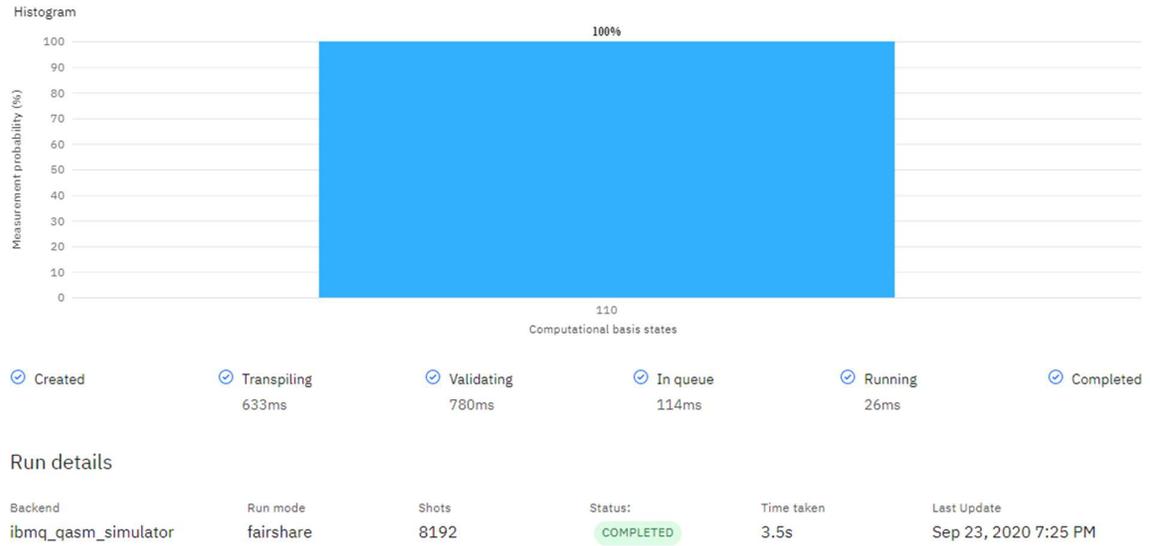

FIG. 22: IBM Q simulator. Upper figure: Histogram (measurement probabilities) in terms of computational basis states, with 100% of probability in $b_2b_1b_0 = 110$, which evidences an absolute coincidence with the metrics of Fig. 13. Lower figure: Run details of simulator execution, with 8192 shots, and fairshare run mode.

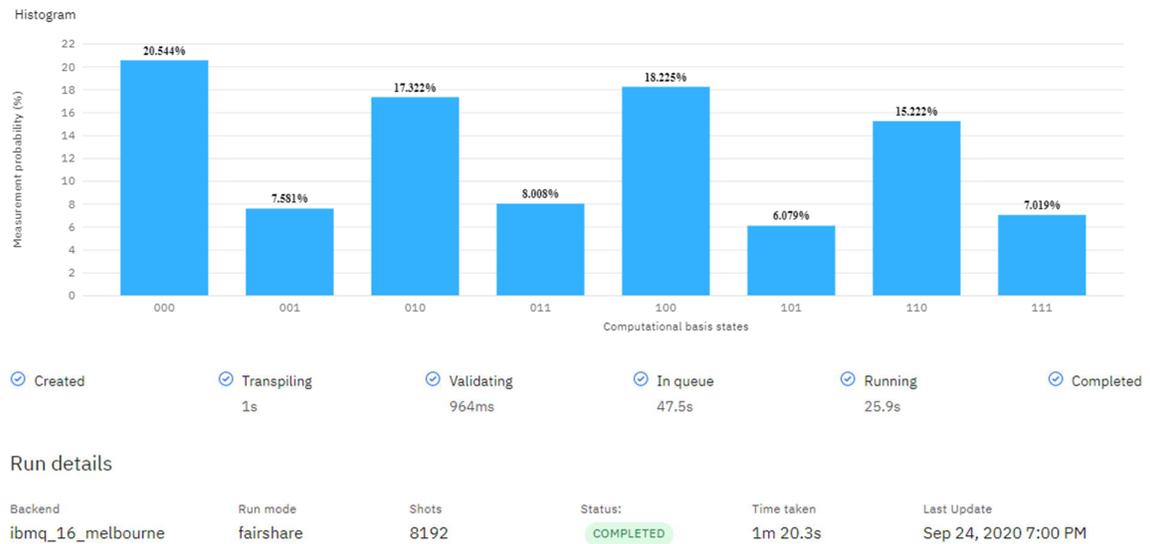

FIG. 23: IBM Q Melbourne processor. Upper figure: Histogram (measurement probabilities) in terms of computational basis states, with (20.544+17.322+18.225+15.222)% = 71.313% of probability in $b_2b_1b_0$ = MM0, and (7.581+8.008+6.079+7.019)% = 28.687% of probability in $b_2b_1b_0$ = MM1, where M is a meta-symbol that represents 0 and 1, at the same time. The difference of almost 29% between the simulator and Melbourne processor shows decoherence of the latter for this experiment. Lower figure: Run details of simulator execution, with 8192 shots, and fairshare run mode.

**Key, and two quantum repeaters:** This configuration is represented with a performance of 100% in Figs. 24 and 25 on Quirk simulator[44], and IBM Q[40], respectively. Therefore, the important thing is to verify the hypothesis about who is responsible for the drop in performance on the Melbourne processor.

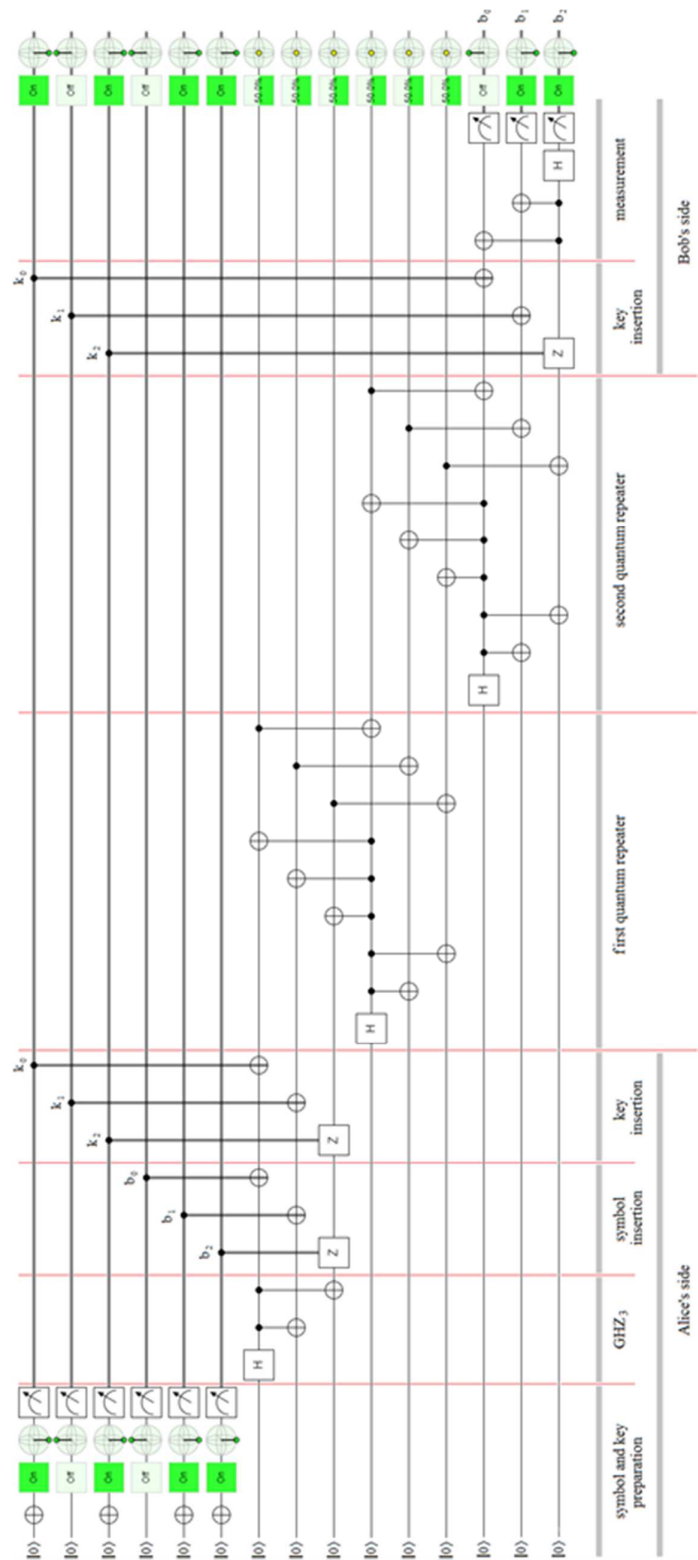

FIG. 24: QKSC implementation on Quirk simulator, with key and two quantum repeaters.

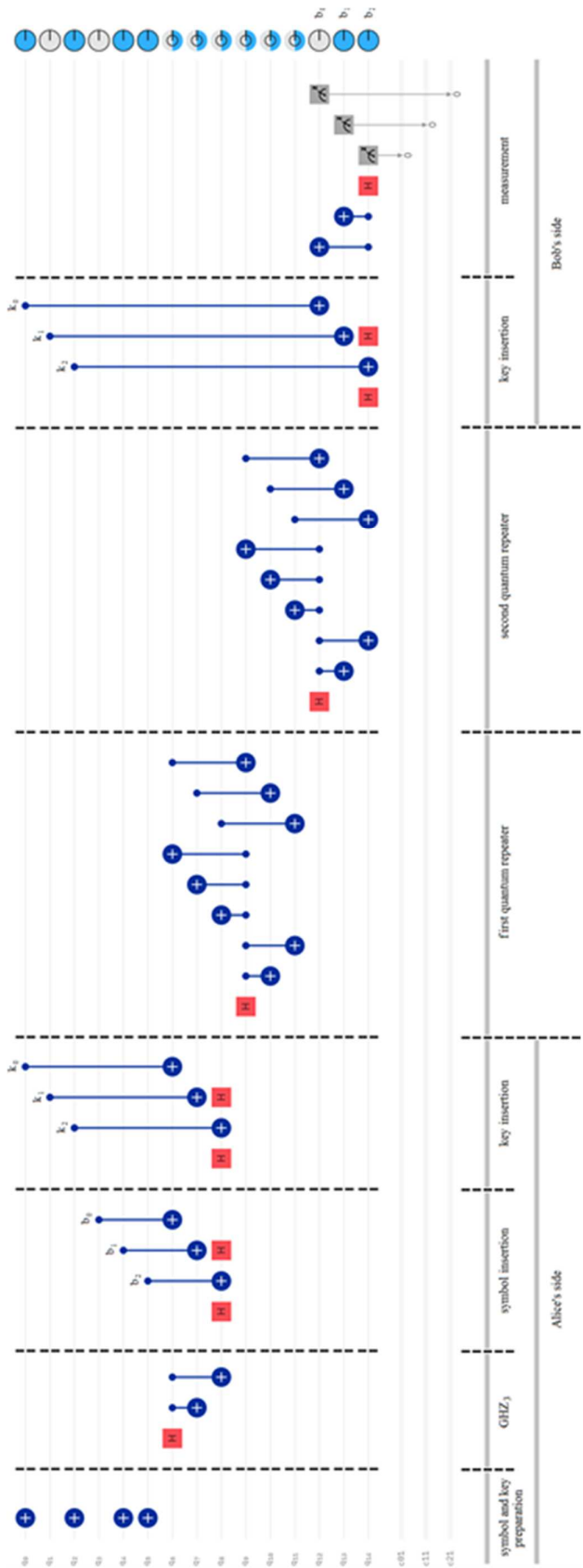

FIG. 25: QKSC implementation on IBM Q, with key and two quantum repeaters.

Figure 26 does not help in the verification of the pending hypothesis because, like Figures 14, 18 and 22, it shows the exact outcomes that would be expected for this case, i.e., with a probability of 100% at $b_2b_1b_0$ = 110. Instead, Fig. 27 shows that the hypothesis is correct given the high level of decoherence in its outcomes. In fact, if the previous experiments were not available and the only information at hand were the outcomes of this figure, an exact conclusion about the performance of the QKSC protocol could not be drawn. However, we must gather all the evidence from the four experiments in an integrative scheme to evaluate the situation more clearly in a comparative way and draw conclusions about who the responsible for the poor performance in the Melbourne IBM Q processor implementation is and why.

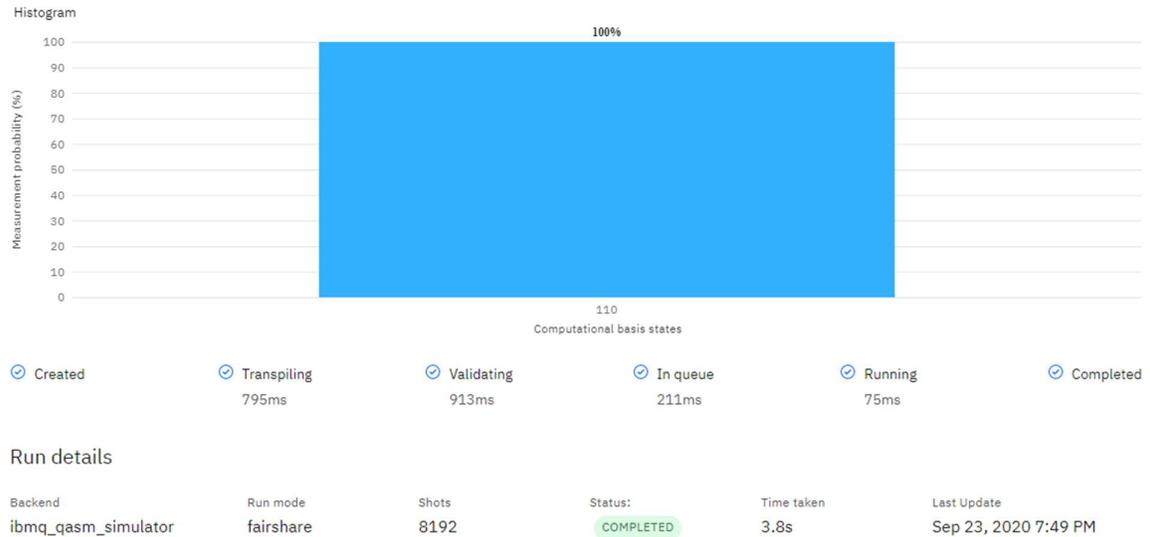

FIG. 26: IBM Q simulator. Upper figure: Histogram (measurement probabilities) in terms of computational basis states, with 100% of probability in $b_2b_1b_0$ = 110, which evidences an absolute coincidence with the metrics of Fig. 13. Lower figure: Run details of simulator execution, with 8192 shots, and fairshare run mode.

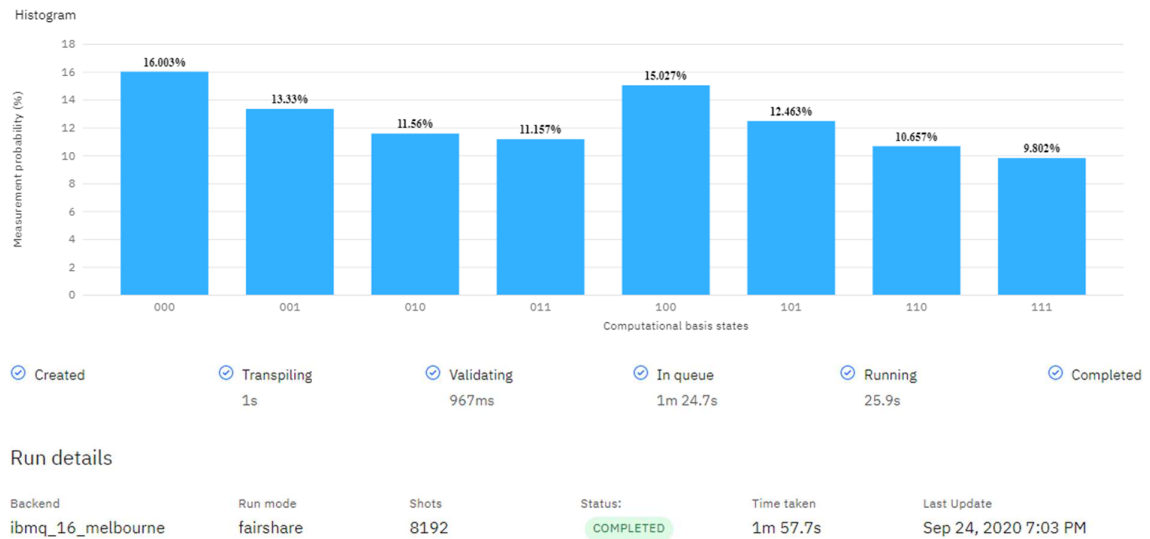

FIG. 27: IBM Q Melbourne processor. Upper figure: Histogram (measurement probabilities) in terms of computational basis states, with (16.003+11.56+15.027+10.657)% = 53.247% of probability in $b_2b_1b_0$ = MM0, and (13.33+11.157+12.463+9.802)% = 46.752% of probability in $b_2b_1b_0$ = MM1, where M is a meta-symbol that represents 0 and 1, at the same time. The difference of almost 47% between the simulator and Melbourne processor shows decoherence of the latter for this experiment. Lower figure: Run details of simulator execution, with 8192 shots, and fairshare run mode.

**Analysis of outcomes:** The four previous experiments are summarized in Table II and Fig. 28, where the outcomes were grouped according to the use or not use of a key, for one or two quantum repeaters, on the IBM Q simulator or Melbourne. In particular, Fig. 28 highlights the errors in red, showing that the responsible for them is the second quantum repeater and not the key. This is mainly due to the fact that each quantum repeater involves a large number of CNOT gates, which introduce bit flip errors to which Melbourne is extremely sensitive.

Table II: Comparative analysis between IBM Q simulator and Melbourne processor of all experiments.

| CBS | keyless, 1 QR | | keyless, 2QRs | | key, 1 QR | | key, 2QRs | |
|---|---|---|---|---|---|---|---|---|
| | simulator | Melbourne | simulator | Melbourne | simulator | Melbourne | simulator | Melbourne |
| 110 | 100 % | 71.241 % | 100 % | 59.497 % | 100 % | 71.313 % | 100 % | 53.247 % |
| rest | 0 % | 28.758 % | 0 % | 40.503 % | 0 % | 28.687 % | 0 % | 46.752 % |

QR: Quantum repeaters, and
CBS Computational basis states.

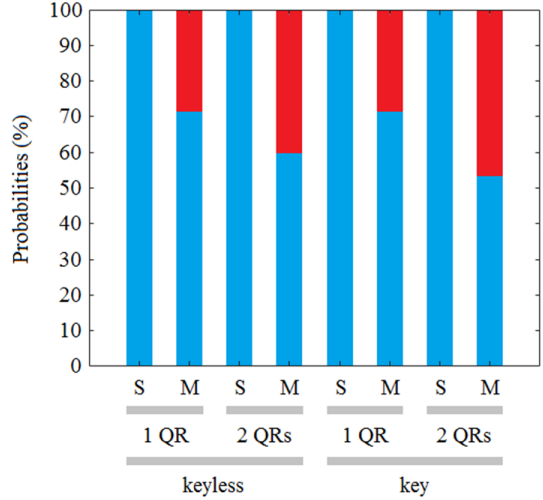

FIG. 28: Probabilities of all experiments performed, with the expected values in blue and errors in red, where S means simulator, M Melbourne, and QR quantum repeater.

**QKD vs QKSC in a real context:** In Fig. 29, we can observe a comparative analysis in the use of QKD and QKSC protocols in a real and conspicuous case[77] of secure quantum communication. Alice (A) and Bob (B) are in two allied submarines, trying to communicate between each other through a quantum satellite. The submarines are submerged and are connected to their respective buoys on the surface, Alice in red and Bob in blue, using resistant cables to the elements of the environment. Eve's not-ally submarine (E) with its respective black buoy is submerged, silent and pre-existing to the arrival of Alice's submarine in the area. All the buoys are tiny in real life but exaggerated in size in Fig. 29 and just barely rise above the surface of the sea. Eve's submarine is far enough away so as not to be detected by Alice's sonar, and close enough to be under the coverage area, in pink in Fig. 29(a), due to the electromagnetic radiation of the classic channels (PCh: Public, and DCh: Data) that Alice and Bob will use to share a key via a QKD protocol and transmit encrypted information using that key. As we have seen in Fig. 3(a), QKD protocols require three channels: Quantum (QCh), public (PCh), and data (DCh) channels in order to distribute a key, and send the encrypted message based on that key. Obviously, if we were to use a QKD-based cryptographic system to communicate to Alice's and Bob's submarines, we would have to resort to a configuration like the one in Fig. 29(a), where the orange rays represent the polarized or entangled photons required by the QKD protocol, while the rays in gray represent the classic PCh and DCh channels. On the other hand, as we mentioned in the first section, the QKD protocols have four loopholes related to the excessive exposure of the key in the channels, which is consistent with the problem posed in Fig. 29(a). Therefore, we must consider an alternative based on the best tools available, using a CubeSat of 6 units developed by Implexum LLC[78], with a low Earth orbit (LEO) at an altitude of 500 km with several minutes of presence between both allied submarines. This platform uses a double telescope to make focus on both allied submarines at the same time with a footprint of 200 m on each one. Based on this platform it is possible to use a cryptographic protocol with fewer channels like QKSC, which results in a simplification of the Cubesat and the buoys associated with each submarine. Thus, the combination established between the aforementioned Cubesat and the QKSC protocol (with a dynamic key like the one explained above) seems to be the ideal solution, although several alternatives arise taking into account the transmission of the first key, and the reduced footprint generated by that Cubesat:

a) the first key is not transmitted, but was pre-agreed before the submarines set sail,
b) the first key is transmitted directly to both allied submarines thanks to a multi-photon source[65] and without any protocol, emphasizing the little footprint,
c) similar to the previous case but using a single-photon source[79-81],
d) the first key is transmitted via a QKD protocol using a multi-photon source[65], emphasizing the little footprint again, or
e) similar to the previous case but using a single-photon source[79-81].

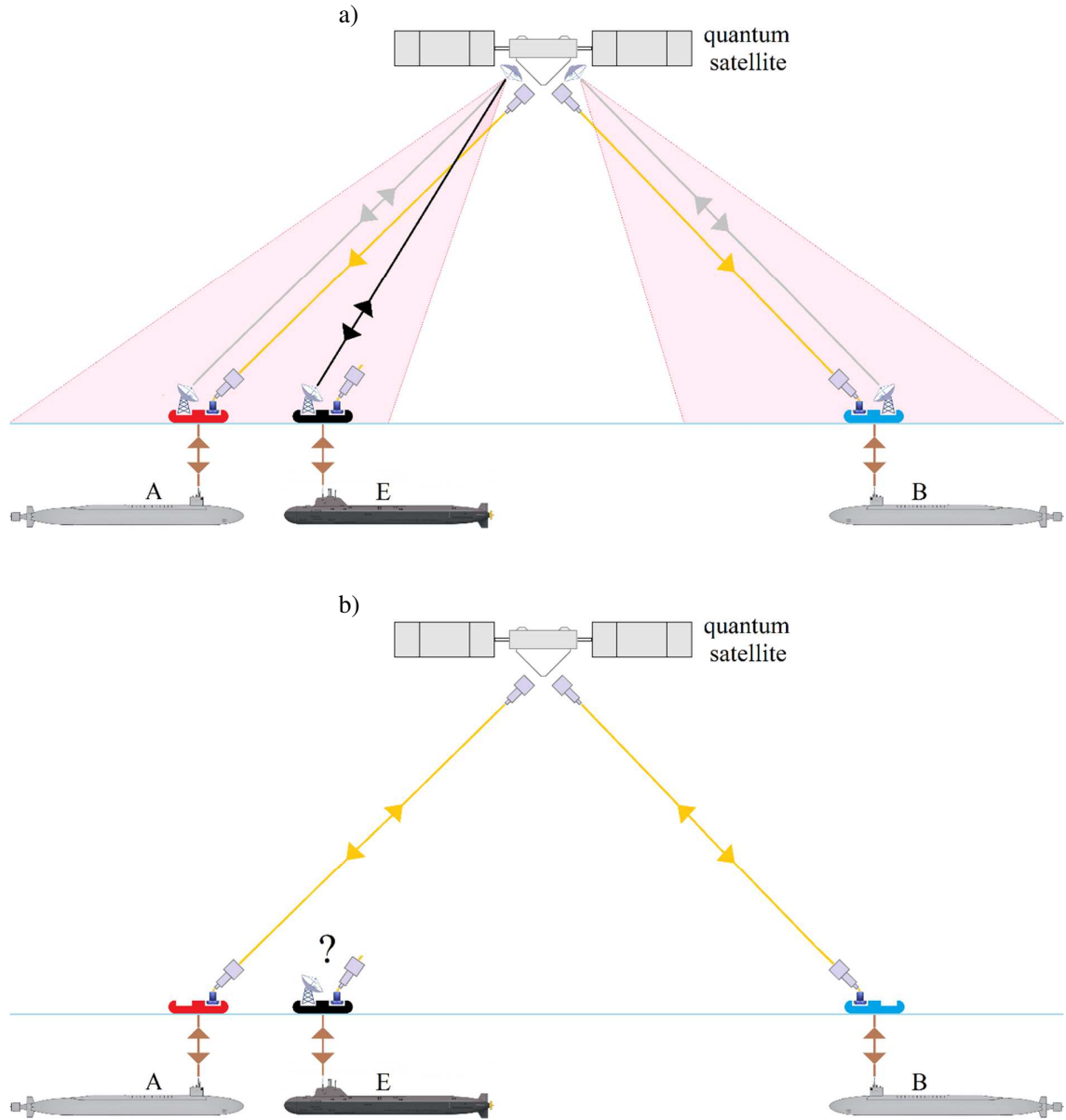

FIG. 29: A quantum satellite to communicate Alice (a red buoy on sea in point A) and Bob (a blue buoy on sea in point B). Besides, there is a third ship, the Eve submarine (E), silent, submerged and pre-existing in the Alice's area. Therefore, two alternatives emerge: a) the not ally submarine is at an appropriate distance from Alice's one, i.e., close enough to be affected by the electro-magnetic shadow associated to the satellite footprint (pink triangular sector), and far enough not to be detected by Alice, thus being able to decode and thus alter the bits of the public and data channels used by the QKD protocol, instead, b) using a satellite with a fully optical channel, i.e., quantum channel (QCh), which can focus exclusively on Alice's buoy for transmission and reception of the entangled photons used by QKSC, the not ally submarine has no chance of altering the communication between Alice and Bob. Otherwise, the orange rays in (a) and (b) represent the entangled photons scattered across the satellite, while the brown rays represent the cables between the buoys and the submarines, which are subjected to great forces of stretching and compression, as well as mechanical degradation due to exposure to the environment. Moreover, in (a), the gray rays represent the electromagnetic links that drives the transmission of the classic bits that Alice needs to rebuild the transmitted keys, via a QKD protocol, and then the message, while the black ray represents the intervention of the not ally submarine in the electromagnetic channel, whereas, in (b) the yellow ray represents the only intervening channel (optical link). They are the Alice's and Bob's buoys that reconstruct the transmitted information and emit them to the submarines. Finally, all the elements of the figures are out of proportion in order to make them more visible.

Based on these tools, we would be in a configuration like that of Fig. 29(b), where the security would be total, since the exposure of the key and the message would be practically null, since it is impossible for Eve's submarine and buoy not to be detected being at a distance of less than half a kilometer from Alice's submarine and buoy.

## 5  Conclusions and future works

### 5.1  Conclusions

There are several versions of Quantum Secure Direct Communication (QSDC)[82-86] based on the Super-dense Coding protocol[45-48]. All of them have a formidable performance of transmission, but with a defect which blocked its use. Said defect is the lack of a key in the presence of an eavesdropper on the channel. Taking these results as a reference, we present here a new protocol called QKSC, which is based on an improvement made over the Super-dense Coding protocol with a dynamic key in order to extend its original performance to an unlimited number of bits to be transmitted and preserving the security in the presence of an eavesdropper in the channel. Among the advantages introduced by the novel, we can highlight the following:

- Independently of the use of an authentication channel, which is common in all protocols of Quantum Cryptography, QKSC only adopts one (QCh) of the three (QCh, PCh, DCh) channels used by a QKD-based cryptographic system,

- this is the first protocol of Quantum Cryptography that is presented with a dynamic key, where it changes with the randomness of the message and the previous key,

- it does not transmit the key, i.e., the key is never exposed as it happens in the QKD protocols,

- QKSC protocol is simple to implement, cheap and a low maintenance solution, compared with any QKD protocol,

- this is a complete communication, and data security system,

- unlike a traditional QSDC system, where part of the data used by the protocol goes back and forth between sender and receiver in order to complete its operation, QKSC is an exclusively forward procedure, i.e., QKSC only uses one of the two branches of the QCh used by a QSDC protocol, however, it is easy to develop a bidirectional version of this protocol based on a practical antecedent such as that of Massa, F., *et al*[87],

- key and message are always protected, which is why a multi-photon source can be used to counteract loss of photons in transmission without exposing security to number-photon splitting attacks[58,59], since the novel depends heavily on the permanent distribution of photons, and

- its use is ideal in command, control, communications, and computers (C4); Intelligence, Surveillance and Reconnaissance (ISR); and Network Operations Center (NOC) systems.

In an optical implementation of the QKSC protocol, if the entire infrastructure was successfully intervened by an eavesdropper, there are would still be the protection via a dynamic key as the last line of defense. In fact, as each message usually occupies several characters or samples, and with each character or sample the keys change, then, during a message the key will change as many times as characters or samples the message has. This causes that any eavesdropper does not have the necessary room nor the time to break a message, in fact, not even a single character or sample. He cannot even perform an easy and efficient off-line intervention, in a batch or quasi-batch modalities, such as statistical analysis, pattern recognition, or some disambiguation procedure.

Therefore, QKSC protocol will be protected even from future attacks by quantum computers with very similar criteria to post-quantum cryptography[34-37]. Finally, given that eventually only the first key should be distributed by any of the methods described in Section 3, and the rest are generated automatically, and independently, although in a synchronized way both on the Alice's (sender) and Bob's (receiver) side, we can say that the dynamic key system of the QKSC protocol resembles a virtual procedure of key redistribution.

If QKSC were used without a key but at a single-photon level, and an eavesdropper present on the channel executed a number-photon splitting attack, Bob could detect the alteration in the transmission and notify Alice to stop all transmission. In this way, the integrity of the message would be preserved, but the transmission between allies would be interrupted. On the other hand, if a multi-photon source is used, the detection of the eavesdropper action by Bob would be compromised. This is the reason for the inclusion of a dynamic key, which would preserve the integrity of the message so that the transmission would never be interrupted even if the presence of an eavesdropper on the channel had been confirmed.

Finally, QKSC is the first protocol of Quantum Cryptography with a complete secure information transmission scheme using a dynamic key.

### 5.2 Future Works

They basically consist of the optical implementation of the QKSC protocol both in the case of lines of optical fibers, as well as in configurations such as those in Fig. 29(b).

**Data Availability**

The experimental data that support the findings of this study are available in ResearchGate with the identifier https://doi.org/10.13140/RG.2.2.17197.72165.

**Acknowledgements**

M. Mastriani thanks to Prof. S.S. Iyengar, director of the School of Computing and Information Sciences of Florida International University for all his help and support.

**Competing interests**

Author declare they has no competing interests.

**Funding**

This research received no external funding.